%% file: demoulin_MV_arXiv.tex
\documentclass[]{aa}
\usepackage{url}
\usepackage{natbib}               
\usepackage{graphicx}             
\usepackage{amssymb}              
\usepackage[usenames,dvipsnames]{color} 
\usepackage{txfonts}

\include{mv1_flux_balance_definitions}

\begin{document}

\title{Exploring the biases of a new method based on minimum variance for interplanetary magnetic clouds}

\titlerunning{Improving minimum variance method for expanding symmetric flux ropes}
\authorrunning{D\'emoulin et al.}

\author{P. D\'emoulin\inst{1}, S. Dasso\inst{2,3} \and M. Janvier\inst{4}}
   \offprints{P. D\'emoulin}
\institute{
$^{1}$ LESIA, Observatoire de Paris, Universit\'e PSL, CNRS, Sorbonne Universit\'e, Univ. Paris Diderot, Sorbonne Paris Cit\'e, 5 place Jules Janssen, 92195 Meudon, France, \email{Pascal.Demoulin@obspm.fr}\\
$^{2}$ CONICET, Universidad de Buenos Aires, Instituto de Astronom\'\i a y F\'\i sica del Espacio, CC. 67, Suc. 28, 1428 Buenos Aires, Argentina, \email{sdasso@iafe.uba.ar} \\
$^{3}$ Universidad de Buenos Aires, Facultad de Ciencias Exactas y Naturales, 
Departamento de Ciencias de la Atm\'osfera y los Oc\'eanos and Departamento 
de F\'\i sica, 1428 Buenos Aires, Argentina, \email{dasso@df.uba.ar}\\
$^{4}$ Institut d'Astrophysique Spatiale, UMR8617, Univ. Paris-Sud-CNRS, Universit\'e Paris-Saclay, B\^atiment 121, 91405 Orsay Cedex, France \email{mjanvier@ias.u-psud.fr}\\
}

   \abstract  
   {
Magnetic clouds (MCs) are twisted magnetic structures ejected from the Sun and probed by in situ instruments. They are typically modeled as flux ropes (FRs). 
}
   {Magnetic field measurements are only available along the 1D spacecraft trajectory.  The determination of the FR global characteristics requires the estimation of the FR axis orientation.
   Among the developed methods, the minimum variance (MV) is the most flexible, and features only a few assumptions.  However, as other methods, MV has biases. We aim to investigate the limits of the method and extend it to a less biased method.
   }
   {We first identified the origin of the biases by testing the MV method on cylindrical and elliptical models with a temporal expansion comparable to the one observed in MCs. Then, we developed an improved MV method to reduce these biases.
   }
   {In contrast with many previous publications we find that the ratio of the MV eigenvalues is not a reliable indicator of the precision of the derived FR axis direction.   Next, we emphasize the importance of the FR boundaries selected since they strongly affect the deduced axis orientation.  We have improved the MV method by imposing that the same amount of azimuthal flux should be present before and after the time of closest approach to the FR axis.  We emphasize the importance of finding simultaneously the FR axis direction and the location of the boundaries corresponding to a balanced magnetic flux, so as to minimize the bias on the deduced FR axis orientation.  This method can also define an inner flux-balanced sub-FR.   We show that the MV results are much less biased when a compromize in size of this sub-FR is achieved.     
   }
   {For weakly asymmetric field temporal profiles, the improved MV provides a very good determination of the FR axis orientation.
   The main remaining bias is moderate (lower than $6 \degree$) and is present mostly on the angle between the flux rope axis and the plane perpendicular to the Sun-Earth direction.
   }

    \keywords{Physical data and processes: magnetic fields, Sun: coronal mass ejections (CMEs), Sun: heliosphere 
    }

   \maketitle

\section{Introduction} 
\label{sect_Introduction}

The Sun release mass and magnetic field in a permanent solar wind, and 
also as transients called coronal mass ejections (CMEs).
Coronal remote white-light observations using coronagraphs 
have shown that the distribution of mass in some CMEs present images 
consistent with twisted structures \citep[\eg ,][]{Krall07,Gopalswamy13,Vourlidas13,Wood17}.
When CMEs are observed in the interplanetary medium they are called 
interplanetary CMEs (ICMEs). The association between CMEs and ICMEs was 
well established since several decades \citep[\eg ,][]{Sheeley85}.
In fact, twisted flux tubes, or flux ropes (FRs), are present also in 
several other systems in the heliosphere, such as the  
Sun atmosphere, the solar wind, and different locations of planetary magnetospheres \citep[\eg ,][]{Fan09,Imber11,Smith17,Pevtsov14,Kilpua17}. 
Flux ropes can store and transport magnetic energy and, because their magnetic field lines can be strongly twisted, FRs can also contain and transport important amounts of magnetic helicity \citep[\eg ,][]{Lynch05,Dasso09b,Sung09,Demoulin16}.

When FRs are observed \insitu\ by a crossing spacecraft, they present a large and coherent rotation of the magnetic field vector.  
A particular set of events that present these FR characteristics corresponds to magnetic clouds (MCs). 
They are also characterized by a stronger magnetic field and a lower proton temperature than the typical solar wind \citep[\eg ,][]{Burlaga81,Gosling90}. Thus, they have a low plasma beta, so that magnetic forces are expected to be dominant.

Interplanetary MCs have been systematically observed from the 80's, and several models have been proposed to describe their magnetic structure.
The simplest one and generally used to model the field in MCs is an axially symmetric cylindrical magneto-static FR solution, with a relaxed linear force-free field, the Lundquist's model \citep{Lundquist50,Goldstein83}.
This model describes relatively well the field distribution for a significant number of observed MCs \citep[\eg ,][]{Burlaga82b, Lepping90, Burlaga95, Burlaga98, Lynch03, Dasso05, Lynch05, Dasso06}. 
Still, the Lundquist solution is known to have difficulties in fitting the magnetic field strength, in particular it was found that it frequently overestimates the axial component of the field near the FR axis \citep[\eg ,][]{Gulisano05}. 

Several other models have been developed.
These models include more general FR properties, such as non-linear force-free field \citep{Farrugia99},
non-force free magnetic configurations \citep[\eg ,][]{Mulligan99, Hidalgo00, Hidalgo02, Cid02, Nieves16}, and 
models with non-cylindrical cross section \citep[\eg ,][]{Vandas03, Nieves09, Nieves18b}. 

In particular, the elliptical model of \citet{Vandas03} provides a better fit to some observed MCs having a field strength more uniform than in the Lundquist solution.  This indicates the existence of some flat FRs \citep{Vandas05}.
On the other hand, from a superposed epoch analysis, \cite{Masias-Meza16} showed that while slow MCs present a symmetric profile of the magnetic field, fast ones present a magnetic intensity profile with a significantly stronger $B$ near the front than at the rear.

The models presented above were typically compared and fitted to MC \insitu\ observations, which are limited to the 1D cut provided by the spacecraft trajectory inside the FR.  This allows a local reconstruction of the FR cross section, 
and then it is possible to make estimations of global magnetohydrodynamic (MHD) quantities, such as magnetic fluxes, twist, helicity, and energy \citep[\eg ,][]{Dasso03, Leamon04, Dasso05, Mandrini05, Qiu07, Demoulin16, Wang16}.  
However, all these estimations depend directly on the FR orientation, and thus on the quality of the method used to get it. Moreover, the orientation itself is an important property of a given FR in order to compare it with its solar origin \citep[\eg ,][]{Nakwacki11,Isavnin13,Palmerio17}.

Different methods to estimate the FR orientation from spacecraft observations have been developed and applied to MCs. 
Some authors have used the so-called Grad-Shafranov method to get the FR orientation, which consists in applying the Grad-Shafranov formalism, valid for describing general MHD magnetostatic equilibria invariant in one direction \citep{Sonnerup96,Hau99,Sonnerup06,Isavnin11}. This method use both the magnetostatic constraints and the folding of the same magnetic field line when it is crossed by the spacecraft during its in-bound and out-bound travel across the FR.
However, the method cannot recover the FR axis for the simplest magnetic configurations, such as symmetric FRs or when the spacecraft crossed the FR close to its axis. \citet{Hu02} argued that observed FRs are typically asymmetric and that this asymmetry removes the above difficulty. In fact a detailed study, D\'emoulin \etal, in preparation, show that for FRs typical of MCs the Grad-Shafranov method provides a biased orientation, which depends on the symmetry properties of the magnetic field components.

Finally, the method that has less hypotheses, mainly because it does not introduce the details of a given model, is the Minimum Variance method (MV). 
The MV method just requires a well ordered large scale variance of $\bf{B}$ in the three spatial directions.  
MV has been extensively used to find the orientation of structures in the interplanetary medium \citep[see \eg ,][]{Sonnerup67, Burlaga82b, Hausman04, Siu-Tapia14}.
Several authors have shown that the MV method estimates quite well the orientation of the FR axis, when the distance between the axis and the spacecraft trajectory in the MC is small with respect to the FR radius
\citep[see \eg ,][]{Klein82, Bothmer98, Farrugia99, Xiao04, Gulisano05, Gulisano07, Ruffenach12, Ruffenach15}. 
Other authors use the MV method to get a first order approximation for the MC orientation.  Then, they use this estimation as a seed in a non-linear least squares fit of a magnetic model to the data.  This approach is expected to improve the cloud orientation \citep[\eg ,][]{Lepping90, Lepping03a, Lepping06, Dasso05}.

The most crucial information to introduce in any of the methods to get the FR orientation is the start and end time of the FR. 
This is a major problem because different authors generally define the FR boundaries at different places/times \citep[\eg ,][]{Riley04c,Russell05,Dasso06,Al-Haddad13}, with the consequent difference on the determination of both the axis orientation and the parameters for the proposed models.
One of the reasons at the origin of these different boundaries is the partial erosion of the FR due to magnetic reconnection. If it happens in the FR front, this creates after the closed FR a 'back' which presents mixed signatures of a FR and stationary solar wind \citep{Dasso06}.
This erosion process also can occur at the FR rear, producing a mixed region before the beginning of the real closed FR \citep{Ruffenach15}. This lack of exact information on the FR boundary can have an important influence on the proper determination of the FR axis direction.
Indeed, comparing different methods show a large dispersion of the estimated FR orientation \citep{Al-Haddad13,Janvier15}.  

 In \sect{Geom} we present a geometrical description of FRs in the solar wind, proposing new angles to determine its orientation, and review the MV method and its application to get the orientation of MCs.
 In \sect{Models} we generate synthetic linear force-free FRs with symmetry of translation along their axis, including the possibility of expansion and cylindrical/elliptical cross-sections, and emulate the observed time-series observed by spacecraft.
 Then, in \sect{Tests} the application of several variants of the MV technique are applied to the synthetic clouds, finding biases produced by different MCs properties. In  \sect{Boundary} a deeper analysis of the bias introduced due to a boundary selection is done, and we proposed a new MV method to minimize this bias. Our conclusions are given in \sect{Conclusion}, and in the Appendix we show the expected coupling between field components when MV is applied.

\section{Geometry of Flux ropes}
\label{sect_Geom}

\subsection{Flux rope axis and frame}
\label{sect_Geom_Axis}

\begin{figure}[t!]    
\centering
\includegraphics[width=0.5\textwidth, clip=]{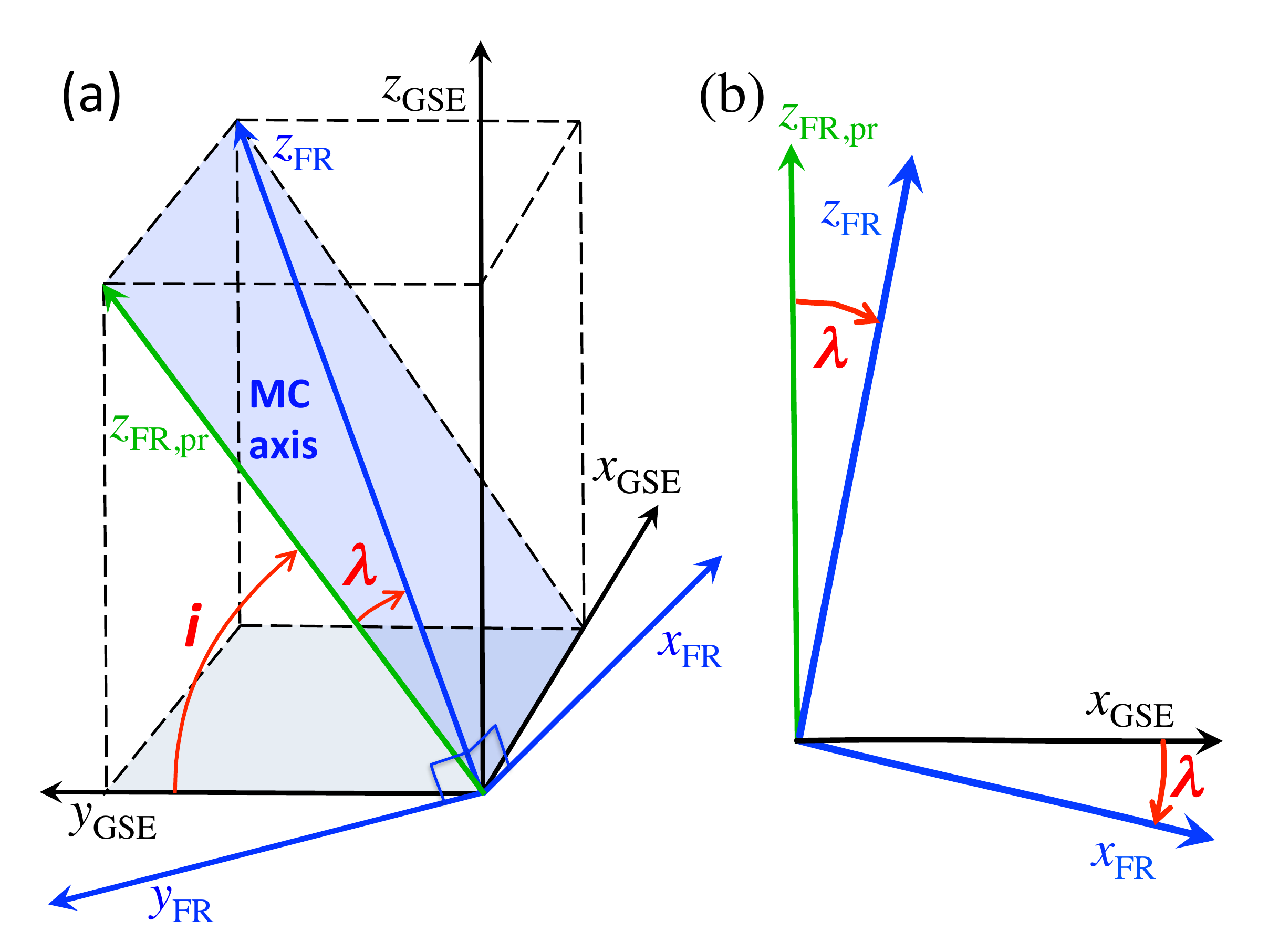}
\caption{Definitions of the angles of the FR axis.
 {\bf (a)} Schema showing the observation, here GSE, and FR frames.  
The FR frame is defined by the vectors $\uvec{x}_{FR}, \uvec{y}_{FR}, \uvec{z}_{FR}$ where $\uvec{z}_{FR}$ is along the FR axis and in the direction of the axial magnetic field, $\uvec{x}_{FR}$ is orthogonal to $\uvec{z}_{FR}$ and in the plane defined by $\uxGSE$ and $\uvec{z}_{FR}$, and $\uvec{y}_{FR}$ completes the orthonormal direct basis. 
Two rotations are needed to pass from the GSE frame to the FR frame.
They are defined by the angles $\iA$ and $\lA$ which are respectively the inclination and the location angle (spherical coordinates with the polar axis $\uxGSE$). 
 {\bf (b)} Schema showing the meaning of the angle $\lA$ in the plane of the FR axis shown in panel (a) in light blue. $\uvec{z}_{pr,FR}$, in green, is the projection of $\uvec{z}_{FR}$ in the plane orthogonal to $\uxGSE$, see panel (a).
}
 \label{fig_schema}
\end{figure}

The coordinates generally used for the analysis of data provided by spacecraft located in the vicinity of Earth are defined in the Geocentric Solar Ecliptic (GSE) 
system of reference. It is defined with an orthogonal base of unit vectors ($\uxGSE, \uyGSE, \uzGSE$). 
$\uxGSE$ points from the Earth toward the Sun, 
$\uyGSE$ is also in the ecliptic plane and in the direction opposite to the Earth rotation motion around the Sun, and $\uzGSE$ points to the north pole of the heliosphere (\fig{schema}a).  A similar coordinate system is the heliographic radial tangential normal (RTN) system of reference \citep[\eg ,][]{Fraenz02} where the following could also be transcripted.

The FR axis orientation is classically defined with respect to the GSE system in spherical coordinates with the polar axis chosen as $+\uzGSE$, 
by two angles: the longitude ($\pA$) and the latitude ($\tA$).  
However these angles are not the natural ones to describe the FR axis orientation with respect to its translation direction ($\approx -\uxGSE$).
  Indeed, the geometrical configuration of the spacecraft crossing is defined both by the closest approach distance and the angle between the spacecraft trajectory and the FR axis orientation.  A rotation of the FR around the spacecraft trajectory does not change the geometry of the spacecraft crossing, while it changes both $\pA$ and $\tA$.  This implies that the same crossing geometry is present along curves defined in the \{$\pA,\tA$\} space.
Even worse, this choice of reference system has the disadvantage to set the polar axis ($\theta = \pm 90$\degree) along $\uzGSE$ which is both a possible and an un-particular axis direction for MCs while the polar axis is singular as it corresponds to any values of $\pA$.  Therefore, the coordinates \{$\pA,\tA$\} are not appropriate to study the orientation of the FR.  

The motion of MCs is mainly radial away from the Sun, especially away from the corona (where limited deflection could occur).  Then, the radial direction is a particular direction. We then set a new spherical coordinate system having its polar axis along $\uxGSE$ (\fig{schema}a).  This direction could correspond to the spacecraft crossing a FR leg.  However, such crossing does not allow to detect the rotation of the magnetic field as the FR is typically only partially crossed on one side. It implies that FRs with an axis direction almost parallel to $\uxGSE$ are mostly not present in MC data sets. Then, the direction $\uxGSE$ can be used as the polar axis of the new spherical coordinate system. 

The local axis direction of the FR is called $\uzFR$ (with $\BzFR >0$ in the central region of the FR).
We define the inclination on the ecliptic ($\iA$) and the location ($\lA$) angles as shown in \fig{schema}.  More precisely, let us define the unit vector $\uvec{z}_{pr,FR}$, in green in \fig{schema}a, along the projection of the FR axis on the plane defined by ($\uyGSE$, $\uzGSE$), then $\iA\ $ is the angle from $\uyGSE$ to $\uvec{z}_{pr,FR}$.  Similarly, $\lA$ is the angle from $\uvec{z}_{pr,FR}$ to $\uvec{z}_{FR}$. 

  If the FR axis is located in a plane, $\iA$ is the inclination of this plane (in light blue) on the ecliptic as shown in \fig{schema}a.
  And if we further suppose that the distance to the Sun is monotonously decreasing from the FR apex to any of the FR leg when following the FR axis, the angle $\lA$ is evolving monotonously along the FR from $-90\degree$ in one of the FR leg, to $\lA=0$ at the apex, then toward $\lA = 90\degree$ in the other leg.
  It implies that $\lA$ can be used implicitly as a proxy of the location where the spacecraft intercepts the FR \citep[as shown in Fig.1c of][]{Janvier13}. 
Finally, the above conclusions extend to other coordinate systems such as RTN. 

We next define the FR frame.  $\uzFR$ corresponds to the FR axis, as defined before.
Since the speed of MCs is much higher than the spacecraft speed, we assume a rectilinear spacecraft trajectory defined by the unit vector $\ud$.  
We define $\uyFR$ in the direction $\uzFR \times \ud$ and $\uxFR$ completes the right-handed orthonormal base ($\uxFR ,\uyFR ,\uzFR $) which defines the FR frame (\fig{schema}a). 

\subsection{Minimum variance technique applied to flux ropes}
\label{sect_Geom_mv}
 
The MV method finds the directions in which the projection of a series of $N$ vectors has an extremum mean quadratic deviation \citep[\eg ,][]{Sonnerup98}.
This method can be applied to the time series of the magnetic field $\vB$ measured \insitu\ across MCs, and it provides an estimation of the FR axis orientation as follows.

The mean quadratic deviation, or variance, of the magnetic field $\vB$ in a given direction defined by the unit vector $\un$ is:
 \BE   \label{eq_sigma}
 \sigma_n^2 = \frac{1}{N}\sum_{k=1}^{N}
                \left( (\vB^k -\langle \vB \rangle) \cdot \un \right) ^2
            = \langle (B_n-\langle B_n \rangle)^2 \rangle
 \EE
where the summation is done on the $N$ data points of the time series and $\langle \vB \rangle$ is the mean magnetic field.

The MV method finds the direction $\un$ where $\sigma_n^2$ is extremum.
The constraint $|\un|=1$ is incorporated with the Lagrange multiplier variational method.  The resulting set of equations in matrix form is
 \BE     \label{eq_sigma2}
 \sigma_n^2 n_j = \sum_{i=1}^{3}m_{ij} \, n_i = \lambda_L \, n_j
 \EE
where $\lambda_L$ is the Lagrange multiplier and 
 \BE    \label{eq_mij}
m_{ij}=\langle B_iB_j
\rangle-\langle B_i\rangle\langle B_j\rangle
 \EE
The matrix $m_{ij}$ is symmetric with real positive eigenvalues: $0 \leqslant \evmin \leqslant \evint \leqslant \evmax$.  Their corresponding orthogonal eigenvectors
($\vec{\uxMV}$, $\vec{\uzMV}$, $\vec{\uyMV}$) are the directions of minimum, intermediate and maximum variation of the magnetic field, respectively. 

The magnetic configuration of MCs is assumed to be a FR with some generic properties which have direct implications on the eigenvalues and eigenvectors found by the MV applied to MC data as follows.  First, the axial field, $\BzFR$,  is stronger on the FR axis and decline to a low value near the boundary. Second, the azimuthal field vanishes on the axis (to have no singular electric current) and typically grows to a magnitude comparable to the axial field strength at the FR periphery.

For trajectories passing close enough from the FR axis, the azimuthal field is mostly in the $\uyFR$ direction. Then, the variance of $\ByFR$ is expected to be about twice the one of $\BzFR$.  The variance of $\BxFR$ is the lowest one. Then the eigenvectors $(\uxMV, \uyMV,\uzMV)$ provide an estimation of $(\uxFR,\uyFR,\uzFR)$, taking into account that the direction of the eigenvectors provided by MV could need a change of sign in order to satisfy the convention defined before for the FR frame.  Next, since $\uyFR$ is perpendicular to $\ud$ (defining the spacecraft trajectory), then the angle between $\uyMV$ and $\ud$ allows to test how close is the MV frame from the FR frame. 
Finally, as the spacecraft trajectory is further from the FR axis, this estimation of the FR frame is expected to be worse.  This is quantitatively tested in Sects.~\ref{sect_Tests}-\ref{sect_Fluctuations} with the models presented in \sect{Models}.


\section{Test models}
\label{sect_Models}
We describe in this section the geometrical aspects of the spacecraft trajectory, and the simulation of measured magnetic field along the trajectory.  
The magnetic models include key ingredients observed or consistent with observations of MC: typical magnetic profiles, expansion, flatness of the FR cross section and asymmetry due to magnetic reconnection. These models are described in the FR frame, with the origin of the reference system set at the FR axis.

\subsection{Geometry of the flux rope crossing}
\label{sect_Mod_Geometry}

During the spacecraft crossing of the FR, its small acceleration has a negligible effect on the measured quantities \citep[\ie , on the time where they are measured,][]{Demoulin08}. Then, we consider a motion with a constant velocity $\Vc$ during the spacecraft crossing.   
Supposing that the FR moves along $-\uxGSE$, the spacecraft position, $\vec{r}_{S/C}(t)$, in the FR frame is:
  \BA
  x(t) &=& \Vc \, t \cos \lA \nonumber \\
  y(t) &=& \yp          \label{eq_spacecraft} \\
  z(t) &=& \Vc \, t \sin \lA \nonumber 
  \EA
where $t$ is the time with $t=0$ set at the time when the spacecraft has the closest distance, $|\yp|$, to the FR axis. Thus, $t<0$ corresponds to the in-bound trajectory (\ie , when the spacecraft is entering the FR going toward its axis) and $t>0$ to the out-bound one (\ie , when the spacecraft is going away from the FR axis).

In order to provide a generic description of a FR crossing we suppose here that the 3D magnetic field evolution, $\vBFR (x,y,z,t)$, is known in the FR frame (with an analytical or a numerical simulation).  
This magnetic field is transformed to the GSE frame by two rotations (with angles $\iFR$ and $\lFR$).
Then, inserting the spacecraft trajectory $\vec{r}_{S/C}(t)$, \eq{spacecraft}, in $\vB_{\rm GSE}(\vec{r})$, limits the description of $\vB$ to the one observed along the spacecraft trajectory, noted $\vBobs (t)$, which is only a function of time.  
After sampling the time uniformly, this time series models the synthetic magnetic field observed by a spacecraft crossing the magnetic field configuration comparable to \insitu\ data obtained in MCs.

The direct MV method, \sect{Geom_mv}, and later on including some additional procedures, \sect{Boundary}, are applied below to the above synthetic field. The estimated angles, $\iMV$ and $\lMV$, are next compared to $\iFR$ and $\lFR$ allowing to test the performance of the MV method on an ensemble of models by scanning the parameter space.

\begin{figure*}[t!]    
\centering
\includegraphics[width=\textwidth,clip=]{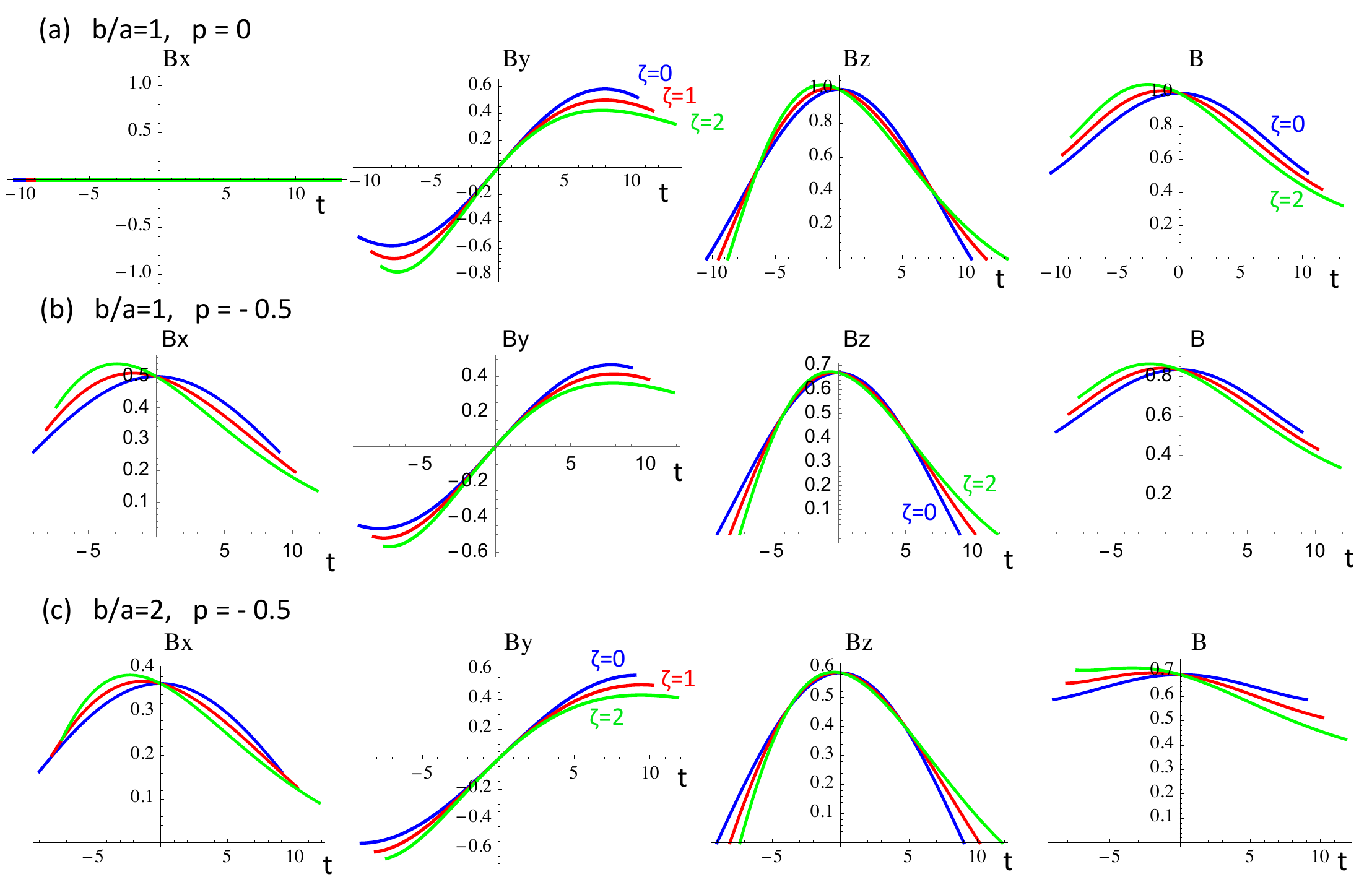}
\caption{Simulations of observed magnetic field components in the FR frame and $|\vB|$ versus time (in hour) for synthetic expanding 
  (a,b) circular and 
  (c) elliptical linear force-free FR with positive magnetic helicity. 
The impact parameter is defined by $p=y/\bo $ where $\bo$ is the FR half-size in the direction orthogonal to the simulated FR crossing.  The FR boundaries are set at $B_z=0$. 
The results with three values of the normalized expansion factor, $\zeta$, are shown with different colors.
A value of $\zeta \approx 1$ is typical according with observations (as deduced from the observed $V_{\rm x,GSE}(t)$ in MCs from 0.3 to 5 AU). The mean observed case, $\bo = 0.1$ AU is shown.  The other parameters, $\Bo =1$ at $t=0$ and $\Vc = 400$ \kms\ are only scaling the axis.
}
 \label{fig_B_components}
\end{figure*}

\subsection{Expanding magnetic field}
  \label{sect_Mod_Expanding}
We consider a general magnetic field configuration, $\vec{\Bo}(x_0,y_0,z_0)$, associated to an element of fluid located at $(x_0,y_0,z_0)$ and defined at time $\to$. 
  Below, to be coherent with \eq{spacecraft}, we simply set $\to=0$ at the time when the FR axis distance to the spacecraft is minimum.
The self-similar isotropic expansion by the factor $e(t)$ implies 
$x = e(t) ~x_0, y = e(t) ~y_0,  z = e(t) ~z_0$ where $x,y,z$ are the new coordinates of the same element of fluid, but at time $t$, with $e(0)=1$. 
 Only an isotropic expansion is considered here since it preserves the force balance of force-free configurations while keeping the magnetic field structure, while a non-isotropic expansion introduces forces which deform the FR so that it would require a numerical simulation to determine the corresponding expanded configuration.   

The expanded magnetic field, $\vB$, is function of space and time:
  \BE \label{eq_B_expansion}
  \vB(x,y,z,t) = \frac{\vB_0~(~x/e(t),~y/e(t),~z/e(t)~)}{e^2(t)} \,.
  \EE
The term $e^2(t)$ at the denominator is included to preserve the magnetic flux. Equation (\ref{eq_B_expansion}) describes the temporal evolution in the 3D space. Introducing \eq{spacecraft} into \eq{B_expansion}, this limits the description of $\vB$ to the observed field along the spacecraft trajectory with $\vBobs (t)$ only function of time.    

The expansion is driven by the decrease of total pressure in the surrounding of the FR as it is moving away from the Sun \citep{Demoulin09}. This total pressure can be approximate{d} by a power law of the distance from the Sun to the FR center ($D(t)$).
The force balance at the FR boundary implies that $e(t)$ is also a power law of $D(t)$: $e(t)=(D(t)/\Do)^\zeta$ where $\zeta$ is the normalized expansion factor \citep[\eg ,][]{Demoulin08} and $\Do$ is a reference distance taken here as the spacecraft distance from the Sun.  
The velocity of the FR core, $\Vc$, is approximately constant during the spacecraft crossing, as in \eq{spacecraft}, then $D(t)=\Vc t +\Do$ with the spacecraft being located at $D=\Do$ at $t=0$.  In this frame work $e(t)$ is:   
  \BE \label{eq_e(t)}
  e(t) = (1+\tau )^\zeta  
       \approx 1+ \zeta ~\tau \,,
  \EE
with
  \BE \label{eq_tau}
   \tau=\Vc ~t~/\Do \,, 
  \EE
which corresponds to the time difference between the observed time and the time when the axis reaches $\Do$, normalized by the travel time from the Sun to the distance $\Do$ at velocity $\Vc$.
The linear approximation on the right side of \eq{e(t)} provides a good approximation for most MCs because $\zeta \approx 1$ and $\tau_{\rm max} <<1$ with $\tau_{\rm max}$ being half of the FR crossing time \citep{Demoulin08}.  
It implies that the expansion typically introduces only a small modification to the observed field profile compared to a case without expansion.

\subsection{Cylindrical flux rope model}
  \label{sect_Mod_Cylindrical}
  
We consider in this subsection a cylindrically symmetric magnetic field configuration for the FR, with its axis along the $z$ direction in the FR frame.  
Then, at a time $t=0$, $\vec{\Bo}(\vec{r})$ is only function of the distance to the axis $\sqrt{x_0^2+y_0^2}$, or equivalently of the relative position of an element of fluid from the FR axis, $\rho$, that is the distance normalized by the FR radius, $\bo$. 
Considering only expansion (\eg , no magnetic diffusion) $\rho$ is preserved during the self-similar isotropic expansion: 
  \BE \label{eq_rho}
  \rho = \frac{\sqrt{x_0^2+y_0^2}}{\bo} = \frac{\sqrt{x(t)^2+y(t)^2}}{b(t)} \,,
  \EE
where $\bo$ and $b(t)$ are the radius of the FR at $t=0$ and $t$ respectively. 
As other spatial variables (\sect{Mod_Expanding}), $b(t)$ is simply related to $\bo$ as $b(t)=e(t)~\bo$.
$\rho$ is a Lagrangian marker of the relative distance to the FR axis.  It follows the magnetic field as the expansion occurs.
In the FR frame, the magnetic field can be written simply as
  \BE \label{eq_B(rho,t)}
  \vB (\rho,t) = \vec{\Bo}(\rho) ~/e^{2}(t) \,.
  \EE
 
The spacecraft rectilinear trajectory satisfies $\yp =$ constant. We define the impact parameter as $p = \yp /\bo$. Using \eq{spacecraft}, \eq{rho} is rewritten along the spacecraft trajectory as:
  \BE \label{eq_rho(tau)}
  \rObs (t) = \frac{\sqrt{(\tau \cos \lA \ \Do/\bo)^2 + p^2}}{e(t)}
  \EE
where the subscript "obs" has been added to specify that $\rho$ is determined along the simulated spacecraft trajectory so $\rObs$ is function of time (and it corresponds to the different elements of fluid observed by the spacecraft), while $\rho$, defined in \eq{rho}, is a general Lagrangian marker, so independent of time. Since the simulated magnetic field along the spacecraft trajectory is $\vec{\Bo}(\rObs (t)) ~/e^{2}(t)$, the magnitudes of $|\tau|$ and $\zeta$ determine how strongly the expansion affects the modeled magnetic field (compared to a case without expansion).

In the FR frame, $\vB = (-\Bt \sin \theta, \Bt \cos \theta, \Bz )$ where $\Bt (\rho)$ and $\Bz (\rho)$ are the azimuthal and axial components respectively.  
Its components along the spacecraft trajectory are:
  \BA
  \BxFR (t) &=&-\frac{\Bt (\rObs)}{e^2(t)} \frac{p}{\rObs ~e(t)}  \nonumber  \\
  \ByFR (t) &=& \frac{\Bt (\rObs)}{e^2(t)} \frac{\Vc ~t \cos \lA}{\bo ~\rObs ~e(t)} 
                                            \label{eq_B_FR} \\
  \BzFR (t) &=& \frac{\Bz (\rObs)}{e^2(t)}  \nonumber 
  \EA

For a linear force-free field (Lundquist's FR):
  \BA
  \BtFR (\rObs) &=& \Bo ~J_1(\alpha \rObs) \nonumber \\
  \BzFR (\rObs) &=& \Bo ~J_0(\alpha \rObs)  \label{eq_B_Lundquist} 
  \EA
where $\Bo$ is the axial field strength of the FR axis. 
We define $\alpha$ as the first zero of the Bessel function $J_0$, $\alpha \sim 2.4$,
then $\rObs =1$ defines the reversal of the axial field component.
Observed MCs have typically a weak axial field component at their boundary so $\rObs \approx 1$, while some cases with $\rObs \gtrapprox 1$ have been observed \citep{Vandas01}.

Examples of simulated magnetic field profiles are shown in \fig{B_components}a,b.    
Observations provide $\zeta \approx 0.8 \pm 0.2$ at 1~AU and $\zeta \approx 0.9 \pm 0.2$ and $\zeta \approx 1.05 \pm 0.3$ in the inner and outer heliosphere, respectively, for unperturbed MCs \citep[not overtaken by a fast stream or by another MC,][]{Demoulin08,Gulisano10,Gulisano12}. Then, we show the cases $\zeta=0,1,2$, which are respectively the cases without, with typical and with twice as large of an expansion rate than observed.  
As $\zeta$ or/and the FR radius increase, the simulated observed field is more asymmetric.  Finally, the case $\zeta=1$ introduces only a moderate asymmetry in all $\vB$ components for not extremely large FR radius (with a radius $ \lesssim 0.1$ AU at 1~AU). 

The $\BxFR$ and $\BzFR$ temporal profiles are nearly symmetric in time while $\ByFR$ is nearly antisymmetric.
A non-zero impact parameter $p$ introduces a component $\BxFR$, whose magnitude increases with  $|p|$, see \eq{B_FR}.  
This property can be used in observations to estimate $|p|$ if a $\Bt (\rho)$ profile is assumed \citep[see Sect. 4.3 in][]{Demoulin09b}. 
Based on different models and the analysis of synthetic FRs, an empirical method to estimate  $|p|$ from the observed $\BxFR$ profile was developed in \cite{Gulisano07}.

\subsection{Elliptical flux rope model}
  \label{sect_Mod_Elliptical}

\citet{Vandas03} generalized the linear-force free field of Lundquist to a FR with an elliptical cross-section.  
Along and across the trajectory, more precisely in the $x$ and $y$ directions of the FR frame, the maximum FR extension is $2~\ao$ and $2~\bo$, respectively. 
The Lundquist's solution is recovered for $\bo=\ao$.
Since the encountered or overtaking solar wind tends to compress the FR in the radial direction away from the Sun, then the relative sizes are expected to satisfy $\ao\leq \bo$.

With invariance along the FR axis, the linear force-free field equations are reduced to 
the Helmholtz equation $\triangle A + \alpha^2 A = 0$ with  $(\Bx,\By,\Bz ) = (\partial A/ \partial y,-\partial A/ \partial x, \alpha A)$.  \citet{Vandas03} solved this equation with elliptic cylindrical coordinates, one of the few coordinate systems where the Helmholtz equation has separable solutions.  
They set the magnetic field to be tangential to the elliptical boundary with a vanishing axial component. For all $\bo/\ao$ values, they found an analytical solution expressed with the even Mathieu function of zero order. 

The above elliptical model can be set in self similar expansion as done above for the Lundquist's field.  The sizes become $a=\ao ~ e(t)$ and $b=\bo ~ e(t)$. The cartesian components of the elliptical model solution $(\Bx,\By,\Bz )$ are first expressed in function of $x,y$. Then, the use of \eqs{spacecraft}{B_expansion} provide the simulated crossing of the FR.  The impact parameter $p$ is generalized to $p = \yp /\bo$, with $\yp$ being the minimum signed distance to the FR axis. Finally, \eq{rho} is generalized to 
  \BE \label{eq_rhoE}
  \rho = \sqrt{(x_0/\ao)^2+(y_0/\bo)^2} = \sqrt{(x/a)^2+(y/b)^2}
  \EE
with $\ao,\bo$ defined at $t=0$ and $a,b$ at $t$.
The selected boundary condition implies $\BzFR (\rho\!=\!1)\!=\!0$.  Finally, along the spacecraft trajectory $\rObs(t)$ is still expressed by \eq{rho(tau)}. 

An example of field component profiles of this elliptical structure is shown in \fig{B_components}c with $b/a=2$.  The main difference with the circular case, \fig{B_components}b, is a flatter $B(t)$ profile. Indeed, since the magnetic tension is less important due to the FR elongation in the $y$ direction, the outward gradient of the magnetic pressure is lower then $B$ is more uniform.
   
\begin{figure*}  
\centering
\includegraphics[width=\textwidth,clip]{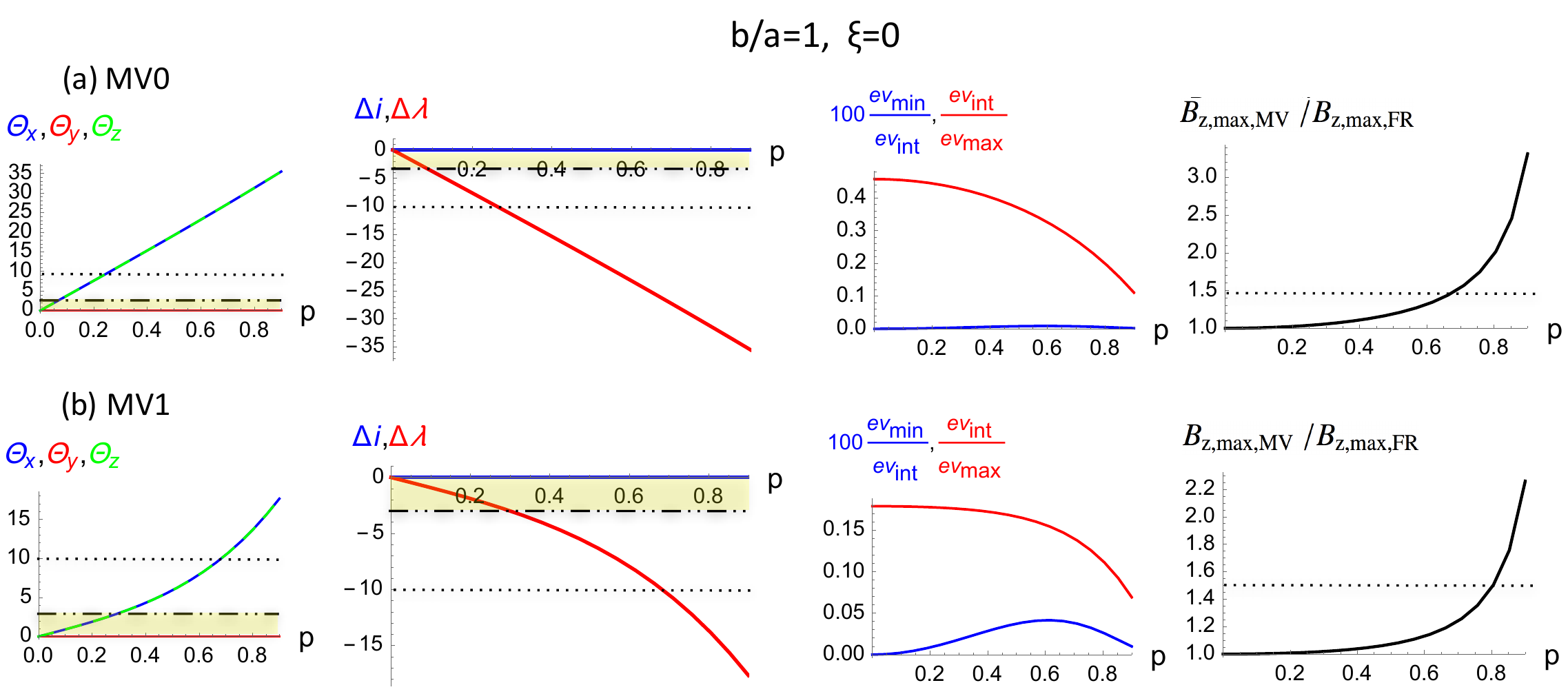}
\caption{Test of the MV method with the Lunquist's FR without expansion ($\zeta = 0$).  The  simulated data are defined within the time interval where $\BzFR \geq 0$.
(a) shows the results for \MV{0} (MV applied to $\vB$) and (b) for \MV{1} (MV applied to $\uvec{B}$). 
The FR axis is oriented with $\iA =90\degree$ and $\lA =0\degree$ so that the simulated observed and FR frames are the same (\fig{schema}). 
\commonCap\  The horizontal dotted lines and the yellow regions are landmarks set at the same locations in different figures for the graphs with the same quantities for a better comparison due to the different scales. 
}
\label{fig_MV_bias_z0}
\end{figure*}

\begin{figure*}[t!]    
\centering
\sidecaption     
\includegraphics[width=0.7\textwidth,clip=]{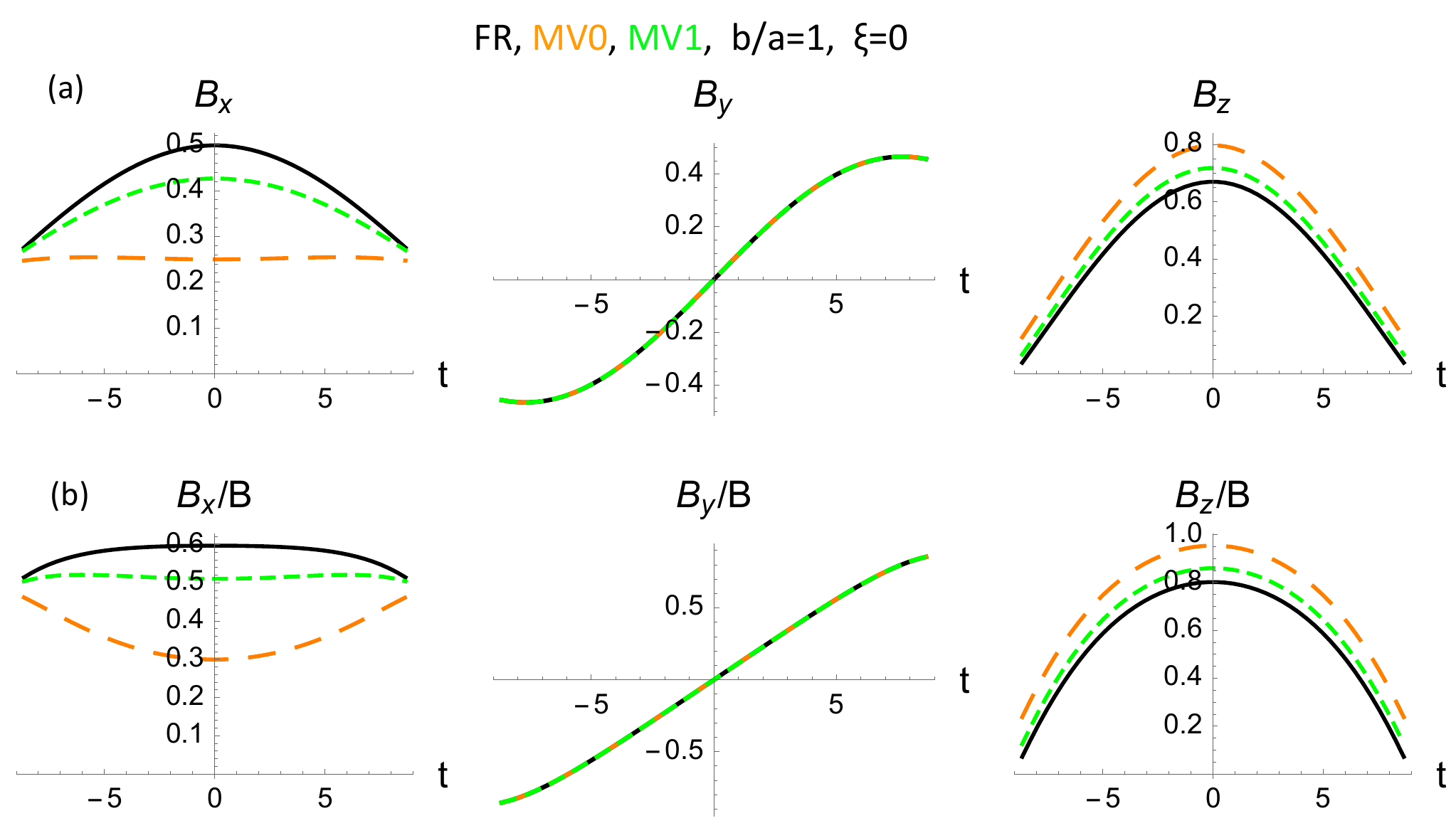}
\caption{Magnetic field components versus time (in hour) in the FR frame (black), in the \MV{0} frame (orange, MV using $\vB$) and in the \MV{1} frame (green, MV using $\uvec{B}$). 
The various components are compared in the same graphs.  All panels are for a static ($\zeta=0$) Lundquist's FR ($b/a=1$).
A relatively large impact parameter, $p=-0.5$, is selected to show the differences ($p<0$ then $\Bx >0$ for this FR with positive magnetic helicity).
(a) shows $\vB$ and (b) shows $\uvec{B}$.  The other parameters are $\Bo =1$ at $t=0$, $\Vc = 400$ \kms, $\bo = 0.1$ AU. 
}
 \label{fig_B_FR_MV_z0}
\end{figure*}

\subsection{Defining flux rope boundaries}
  \label{sect_Mod_Boundaries}

In the absence of magnetic diffusion and reconnection a given value of $\rho$ (\eq{rho}) traces a selected cylindrical shell within the FR.   Along the spacecraft trajectory the front and rear boundaries are located at $\rF$ and $\rR$, respectively.
Below, we consider eroded FRs, with $0< \rF \leq 1$ and $0< \rR \leq 1$, in order to include possible events where the peeling of the FR via magnetic reconnection can be present \citep[see, \eg ,][]{Dasso06, Ruffenach12, Ruffenach15}.  

The temporal series of $\vB$ is computed in the time interval $[\tF,\tR]$ from \eq{B_FR}, or its equivalent for the elliptical case. Because the closest approach is set at $t=0$, \eq{spacecraft}, one has the ordering: $\tF<0<\tR$.   $\tF$ and $\tR$ are found by solving \eq{rho(tau)}, with $\rObs (t)=\rF$ and $\rR$, respectively. We suppose below that $p<\rF$ and $p<\rR$, so that the FR is crossed by the spacecraft. 

The case of the front boundary and $\zeta \geq 0$ (expanding FR) is simple.
$\rObs (t)$ in the in-bound branch is a decreasing function of time (for $t<0$ in \eq{rho(tau)}, the numerator is a decreasing function of time while the denominator is always increasing with time).  This implies that $\rObs (\tF) = \rF $ has a unique solution when $p<\rF$.

The case of the rear boundary is more case-dependent.  This is because both the numerator and the denominator of \eq{rho(tau)} are both increasing with time for $\zeta >0$ and $t >0$.  
For $\zeta<1$, the numerator is growing faster than the denominator so $\rObs (\tR) = \rR $ has a unique solution  (when $p<\rR$).  For larger $\zeta$ values there are two $\tR$ solutions or even none for $\zeta \gtrsim 4 $.  The physical solution in the first case is the one closer to the axis while in the second case the expansion rate is so large that the simulated spacecraft never crosses the rear boundary. 
This last case is not observed as the maximum $\zeta$ values observed within unperturbed MCs is $1.5$ in the inner heliosphere, $1.1$ at 1~AU, and $1.7$ in the outer heliosphere \citep{Demoulin08,Gulisano10,Gulisano12}.

\section{Tests of minimum variance}
\label{sect_Tests}

\subsection{Main procedure}
  \label{sect_Tests_Main}
  
A classical test of the quality of the MV method to distinguish directions 
is to analyze the separation between the variance (eigenvalues) associated with the three eigenvalues $\evmin$, $\evint$ and $\evmax$ of the matrix $m_{ij}$ defined by \eq{mij} (see references in \sect{Introduction}).  Indeed, if two directions have similar variances, so similar eigenvalues, say for example, $x$ and $z$ directions, a rotation of the frame around the third eigenvector ($y$) is expected to provide also similar variances in the two rotated directions ($x'$ and $z'$). This implies that the $x,z$ directions are not well defined.
   
However, a large separation of the three eigenvalues is a necessary but not sufficient condition to have a precise determination of the FR axis direction. A systematic bias could be present due to the mixing of the field components to achieve extremum variances in the eigenvector directions.
To understand this possible bias, let us analyze the Lundquist's FR with a non-negligible impact parameter $p$ as an example. $|\BxFR |$ has a shape comparable to $\BzFR$ (\fig{B_components}b). Then, the MV combines $\BxFR$ and $\BzFR$ by a rotation of frame to produce the flattest possible $\BxMV$ (so with the lowest variance).  This produces a bias in the MV frame orientation which increases with $|p|$, confirming the results of \citet{Gulisano07}.  Indeed, in general, the MV method will mix components which are correlated, see Appendix \ref{sect_MV_sym}.

 More generally, this bias and other ones are present for the axis direction obtained when the MV is applied to expanding FRs.  Below, we first identify, then correct, these biases as much as possible.  
 They are measured by the positive and acute angles $\tx,\ty,\tz$ between the x, y, z directions of the MV and FR frames, respectively (these angles can be defined as signed,  but the absolute value is sufficient for our purpose).  
 Next, we compute $\diA = \iMV-\iFR$ and $\dlA = \lMV-\lFR$ which quantify the difference of axis orientations found by the MV and the correct one simulated with the FR model. 
 Finally, we quantify the bias on the axial field by the ratio $\Bzratio$.
  
From the tests made, we conclude that a large separation of the three eigenvalues is needed but this is far from being sufficient to have a precise axis determination.  For example, for $|p|\lesssim 0.6$, we show in \fig{MV_bias_z0}b a case where the two lowest eigenvalues are more separated when $|p|$ increases, up to $|p|\leq 0.6$, while the bias in the axis orientation increases ($\tx, \tz, \dlA$ in the left panels).     

\begin{figure*}  
\centering
\includegraphics[width=\textwidth,clip]{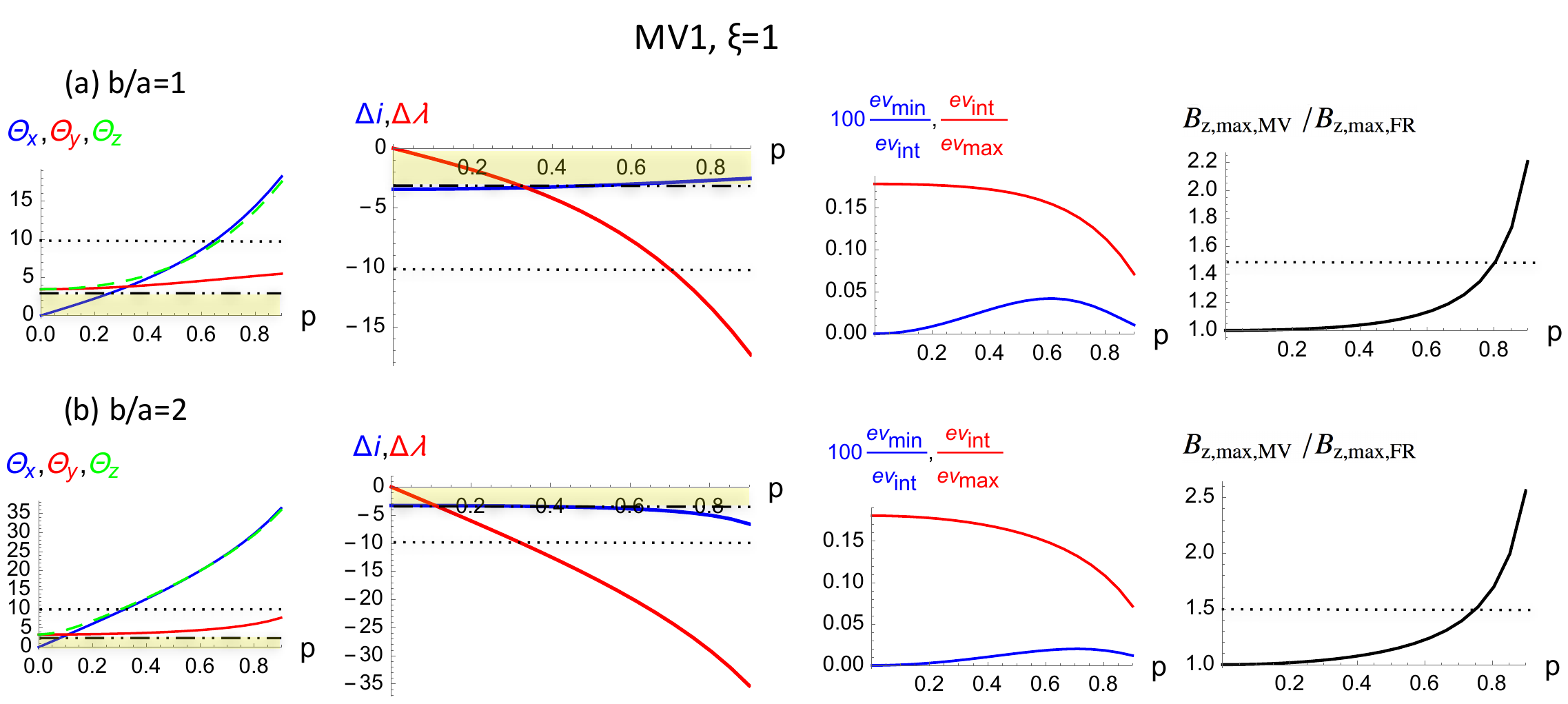}
\caption{Test of the \MV{1} method with FRs expanding with a typically observed rate ($\zeta = 1$).  The  simulated data are defined within the time interval where $\BzFR \geq 0$.
(a) shows the results for a circular FR and (b) for an elliptical one.
The FR axis is oriented with $\iA =90\degree$ and $\lA =0\degree$ so that the simulated observed and FR frames are the same (\fig{schema}).
\commonCap\ The horizontal dotted lines and the yellow regions are landmarks to compare the graphs. 
}
\label{fig_MV_bias_z1}
\end{figure*}

\begin{figure*}[t!]    
\centering
\sidecaption     
\includegraphics[width=0.7\textwidth,clip=]{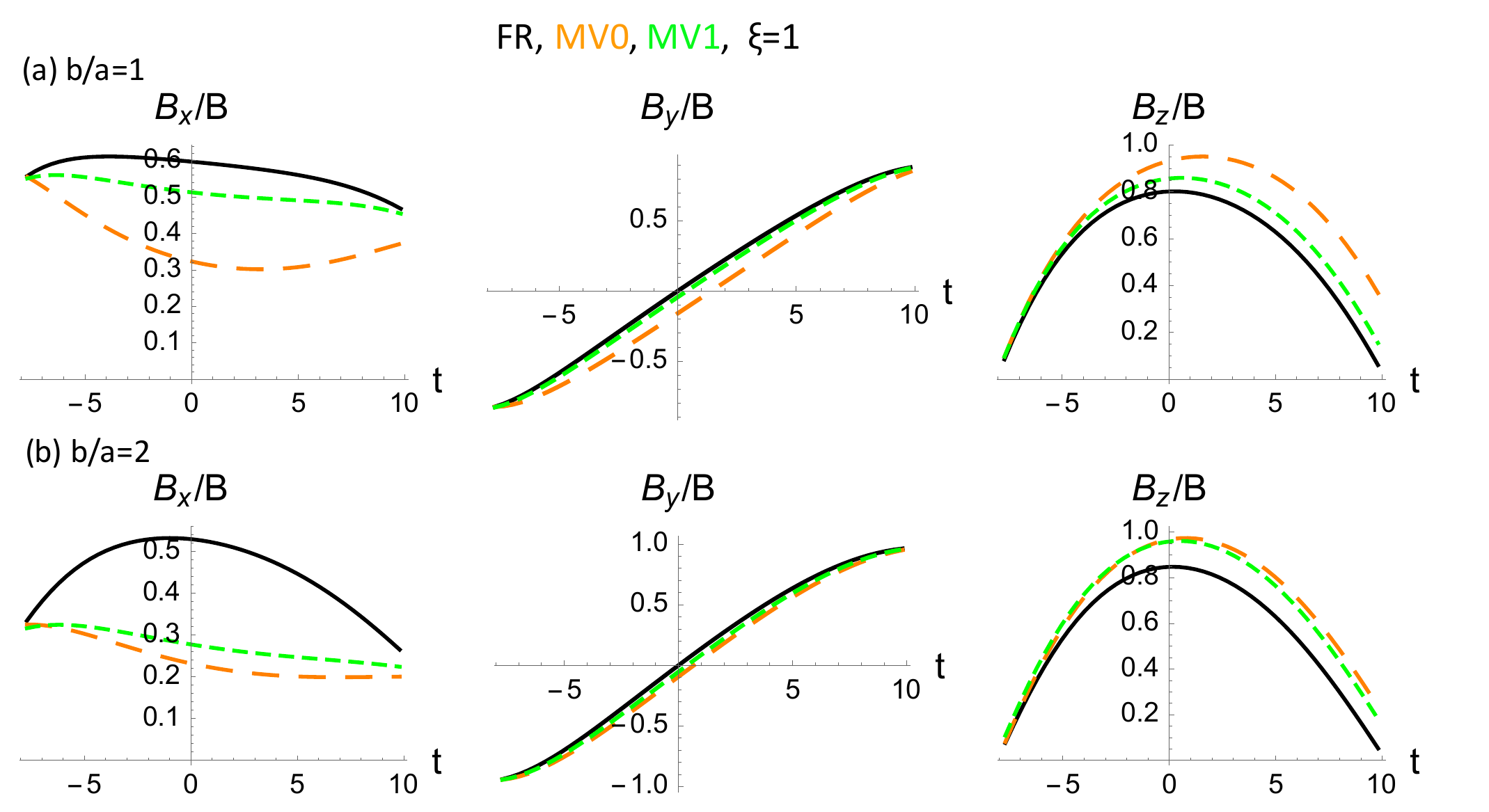}
\caption{Magnetic field components versus time (in hour) in the FR frame (black), in the \MV{0} frame (orange, MV using $\vB$) and in the \MV{1} frame (green, MV using $\uvec{B}$) for FRs expanding with a typically observed rate ($\zeta = 1$). 
A relatively large impact parameter, $p=-0.5$, is selected to show the differences.
(a) shows the results for a circular FR and (b) for an elliptical one.
The other parameters are $\Bo =1$ at $t=0$, $\Vc = 400$ \kms, $\bo = 0.1$ AU. 
}
 \label{fig_B_FR_MV_z1}
\end{figure*}

\begin{figure*}  
\centering
\includegraphics[width=\textwidth,clip]{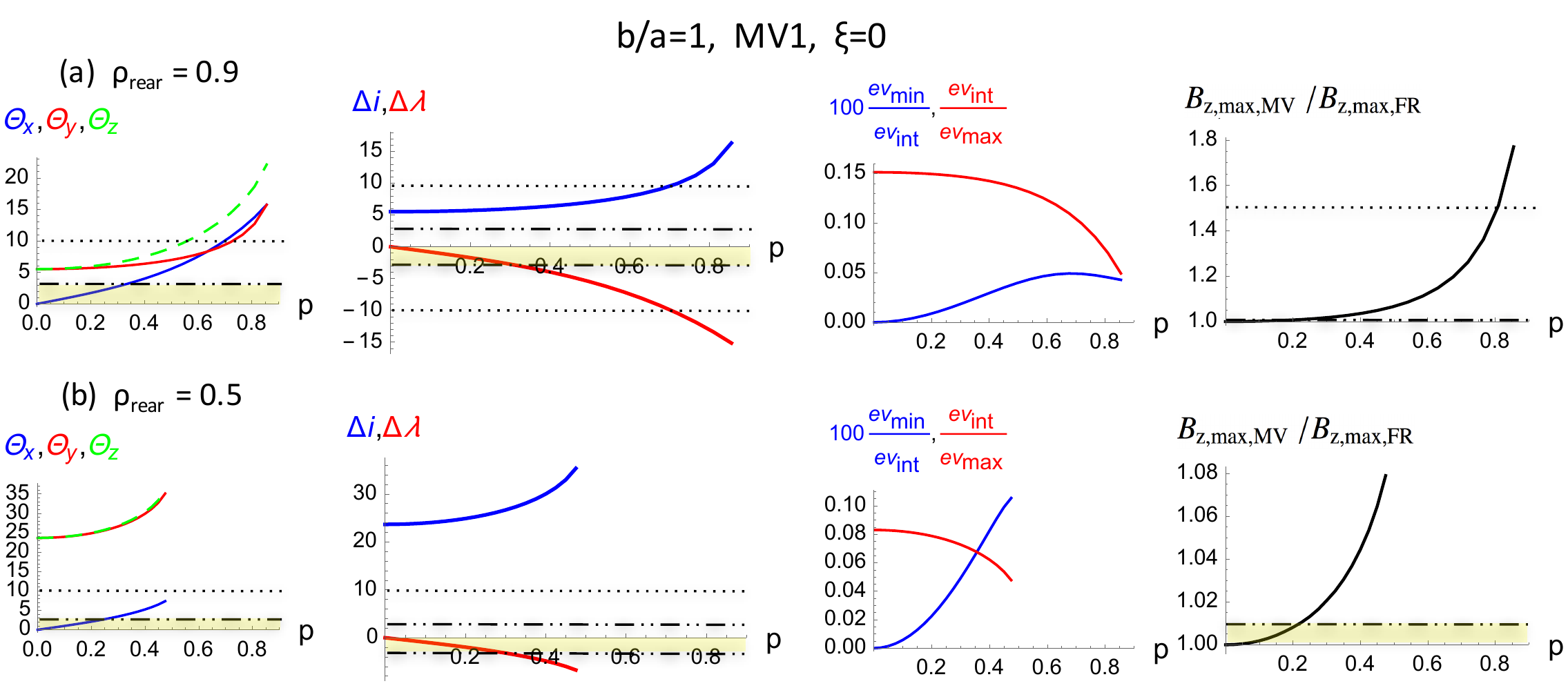}
\caption{Effect of the flux unbalance on the results of \MV{1} applied to the Lundquist's FR without expansion ($\zeta = 0$).  The results are shown for two cases of flux unbalance between the FR front and rear.  The front limit is located where $\BzFR =0$ ($\rF=1$) and the rear limit is set at a fraction $\rR$ of the FR radius.  For comparison the case $\rR =1$ is shown in \fig{MV_bias_z0}b. 
\commonCap\  The horizontal dotted lines and the yellow regions are landmarks. 
 }
\label{fig_rhoR}
\end{figure*}

\subsection{Minimum variance applied to $\vB$ or $\uvec{B}$}
  \label{sect_Tests_B}

The MV technique, described in \sect{Geom_mv}, can be applied to the time series of $\vB$ or $\uvec{B}=\vB/|\vB|$ defined within the modeled FR (and later on with \insitu\ data).  
These approaches are called \MV{0} and \MV{1} and were used, for example, by \citet{Siscoe72} and \citet{Gulisano07}, respectively.

\MV{0}, applied to the Lundquist's model has an orientation bias increasing linearly with $p$ (\fig{MV_bias_z0}a, see $\tx,\tz$ and $\dlA$ variations).  This bias is reduced with \MV{1} (\fig{MV_bias_z0}b). Next, $\ty = \diA$ = 0 for this simple test because $\BxFR ,\BzFR$ are symmetric and $\ByFR$ is antisymmetric with time so that they cannot be mixed by MV (see Appendix \ref{sect_MV_sym}).   
These tests agree with the results of \citet{Gulisano07} and precise the origin of the bias since the use of $\iA$ and $\lA$ is more adapted to separate the relevant magnetic components than the longitude and latitude of the FR axis.
This further justifies the choice of $\iA$ and $\lA$ angles because, when using them, the bias of the orientation remains mainly on $\lA$.

Moreover, the origin of the biases is illustrated for $\zeta=0$ with the $\vB$ components of the Lundquist's FR drawn in the FR frame (black lines) and \MV{0} frame (orange lines) in \fig{B_FR_MV_z0}a. $\BxFR$ and $\BzFR$ are combined to provide a flat $\BxMV$ (\fig{B_FR_MV_z0}a, left panel).
More generally, for a cylindrically symmetric FR, $\Bx =\Bt (\rObs) ~p/\rObs$ in the FR frame from \eq{B_FR}, then $\Bx =$ constant if the azimuthal component $\Bt$ is linear with the radius $\rObs$. In this case, the variance of $\Bx$ vanishes and the \MV{0} method would exactly associate this minimum variance direction to $\uxFR$. In brief, there would be no orientation biases.  However, in general this linear dependence is only approximately true close to the FR center for cylindrically symmetric FR (using a Taylor expansion of $\Bt$). 

  The above bias decreases when normalizing $\vB$ to unity since $|\vB|$ decreases away from the FR axis, compensating partly the decrease of $\Bt$.  It implies that $\Bt /|\vB|$ is more linear with $\rObs$ than $\Bt$.  This effect is directly seen on the $\By$ component for $p=0$ since $\Bt=|\By|$ and it is still well present for $p=-0.5$ when comparing the central panels of \fig{B_FR_MV_z0}. Consequently, $\Bx /|\vB|$ is more uniform than $\Bx$ and the \MV{1} frame is closer to the FR frame than the \MV{0} one (\fig{MV_bias_z0}). Then, the $\vB$ components in the MV1 frame (green lines in \fig{B_FR_MV_z0}) are closer to the ones in the FR frame (black lines) than when \MV{0} is applied to $\vB$ of the Lundquist's field (orange lines).

\subsection{Sign and magnitude of biases}
  \label{sect_Tests_biases}

 The MV bias identified above is due to a mix between $\BxFR$ and $\BzFR$.
By definition of the orientation of the $z$ axis, $\BzFR \geq 0$ in the FR core.   From \eq{B_FR}, $\BxFR$ has the sign of $-p~\Bt$. Then, the bias $\dlA$ has also the sign of $-p~\Bt$. For example, \fig{MV_bias_z0} shows a case with positive helicity, $\Bt \Bz \geq 0$, and $p\geq 0$ so that $\dlA \leq 0$. Other cases are simply obtained by changing the sign of $\dlA$ accordingly to the signs of $p$ and magnetic helicity.
 
The effect of expansion for a typical $\zeta$ value, $\approx 1$, introduces only a small extra bias on $\iA$, $|\diA|<4 \degree$, as seen by comparing \fig{MV_bias_z1}a to \fig{MV_bias_z0}b.  The expansion induces an asymmetry in the field components (\fig{B_components}), then the MV method combines a part of $\ByFR$ with $\BzFR$, \fig{B_FR_MV_z1}a, introducing a bias for $\ty$ and $\diA$ values.  Other quantities ($\tx ,\tz ,\dlA$, the ratio of eigenvalues and $B_{\rm z,max,MV}$) are not significantly affected by the expansion. This small effect of expansion is a generic result for all the models tested.

Increasing the aspect ratio $b/a$ enhances more significantly the bias (\fig{MV_bias_z1}b, we note the change of vertical scale on the left panels).  In fact, since the $B(t)$ profile gets flatter with increasing $b/a$ value (\fig{B_components}c, right panel) \MV{1} is converging to \MV{0} as $b/a$ is larger, so the bias increases (as shown above with \fig{MV_bias_z0} comparing \MV{0} to \MV{1}). In contrast, as $b/a$ increases, the ratio $\evmin /\evint$ decreases (\fig{MV_bias_z1}, compare middle right panels), so this ratio is again not a reliable indicator of the quality of the axis direction determination.  
Furthermore, as $b/a$ increases, $\BxFR (t)$ behaves closer to $\BzFR (t)$ then both \MV{0} and \MV{1} rotate the frame combining $\BzFR$ and $\BxFR$, trying to get $\BxMV$ as flat as possible (compare the two left panels of \fig{B_FR_MV_z1}). This further increases $|\dlA|$ so the bias in the determined axis orientation (\fig{MV_bias_z1}b).

The bias on the axis orientation introduced by the MV implies a systematic overestimation of the axial component $\Bz$, as shown in \fig{B_FR_MV_z0}, which increases with $|p|$ (\fig{MV_bias_z0}, right panels).  This effect is stronger for \MV{0} than \MV{1}  as expected from the orientation bias results.  This bias  on $\Bz$ also increases slightly with $b/a$ value (\fig{MV_bias_z1}, right panels). This is illustrated with an example shown in the right panels of \fig{B_FR_MV_z1}.  The MV method rotates the MV frame so that the bump shape of $\BxFR$ is removed in $\BxMV$. This strengthens $\BzMV$ independently of the sign of $p$ and of magnetic helicity.

\section{Biases due to boundary selection}
\label{sect_Boundary}

\subsection{Defining the flux rope boundaries}
  \label{sect_Boundary_Def}

The boundaries of the time interval selected in the \insitu\ data to find the FR orientation are typically set where abrupt changes of the magnetic field and plasma parameters are detected. However, there are frequently some ambiguities in setting them so that the selected  boundaries depend on the author's specific criteria.  For example, the MC observed  on October 18-20 1995 by the Wind spacecraft was studied by several groups of authors which set different MC boundaries \citep{Lepping97, Larson97, Janoo98, Collier01, Hidalgo02, Dasso06}.

The main origin of these discrepancies is that the measured magnetic field components and the plasma data, such as proton temperature, plasma composition and properties of high energy particles, do not always agree on the extension of the MC, and more generally of the magnetic ejecta \citep[see, \eg ,][]{Russell05}.   A part of this ambiguity is due to the partial magnetic reconnection of the ejected solar FR with the magnetic field of the encountered solar wind or of an overtaking stream.  This reconnection does not affect the field and plasma on the same time scale.  For example a change of magnetic connectivity is affecting on short time scales (on the order of few tens of minutes) the propagation of high energy particles while the mix up of MC and solar wind protons from the distribution core takes days to be mixed after reconnection occurred.  The above processes imply that different boundaries are often defined in different studies of the same MC \citep[\eg ,][]{Riley04c,Al-Haddad13}.  

The selection of boundaries have large implications on the derived FR orientation \citep[\eg ,][]{Dasso06}.  For example, the axis orientation derived from MV and a least square fit to a Lundquist's model can be significantly different and without coherence along the axis of a MC observed by four spacecraft (ACE, STEREO A and B, Wind) for a set of boundaries \citep{Farrugia11}, while consistent results are obtained between both methods and along the FR axis when refined time intervals are used \citep{Ruffenach12}.  
The important effect of the boundaries is further shown in \citet{Janvier15} by the difference in orientations found with the same set of MCs analyzed by \citet{Lynch05} and \citet{Lepping10}, while the FR axes were determined in both cases by the same method (fit of $\vB$ data to Lundquist's model). 
  
\subsection{Testing the effect of boundary location}
  \label{sect_Boundary_Testing}

The effect of the boundary selection on the determined axis is illustrated in  \fig{rhoR} using \MV{1} on a Lundquist's FR without expansion (in order to focus the analysis only on the boundary effects).  
We set the front at $\rF=1$ (where $\BzFR =0$) and allows $\rR <1$. The FR is crossed only for $|p|<\rR$ so the large values of $\diA$, $\dlA$ and $\Bzratio$ obtained for large $|p|$ values 
in \fig{MV_bias_z0}b are absent in figures with $\rR <1$ and the comparison of biases should be done for the same $p$ value.

 Already with $\rR =0.9$ (\fig{rhoR}a), so only with 10\% in radius removed at the FR rear, there is a clear increase of \MV{1} bias (compare to \fig{MV_bias_z0}b where $\rR =1$ is the only different parameter).  $\ty$, which was null in \fig{MV_bias_z0}b, is comparable to $\tx$ and $\tz$ (left panel of \fig{rhoR}a). This introduces mainly a rotation of the estimated FR axis by changing its inclination on the ecliptic plane ($\diA >0$). There is a mix up of $\ByFR$ with the other components in order to increase the variance of $\ByMV$.  This is possible as the components are no longer antisymmetric or symmetric with respect to the closest approach time, then they are partly correlated (see Appendix \ref{sect_MV_sym}).  

As $\rR$ is further decreased the bias on $\iA$ increases, and dominates the bias on $\lA$ for all $p$ values. 
For example, with $\rR =0.5$, so with half of the FR rear removed, this introduces the main bias on $i$ with a rotation of the determined FR axis on the order of $24 -35 \degree$ compared to the known axis direction (\fig{rhoR}b, middle left panel). Even worse, this bias on $i$ is also present for low impact parameter cases (bias $\approx 24\degree$).  This increasing bias is associated with an increasing separation of the two lowest eigenvalues (\fig{rhoR}b, middle right panel). It shows once more that the eigenvalue separation is not a good criteria for estimating the precision of the axis direction defined by MV. 

The location of the FR boundaries also introduces biases on other parameters of the FR such as its radius (by changing the geometry of the crossing) or its axial field strength (by mixing the field components). Still, it has only a moderate effect on $\Bzratio$ since \MV{1} mixes $\ByFR$ with $\BzFR$ and $\ByFR \approx 0$ where $\BzFR$ is maximum (\fig{B_FR_MV_z1}), so that $\Bzratio$ is comparable to the case $\rR =1$ for the same $p$ value (compare the left panel of \fig{MV_bias_z0} to those of \fig{rhoR}).

\begin{figure*}  
\centering
\includegraphics[width=\textwidth,clip]{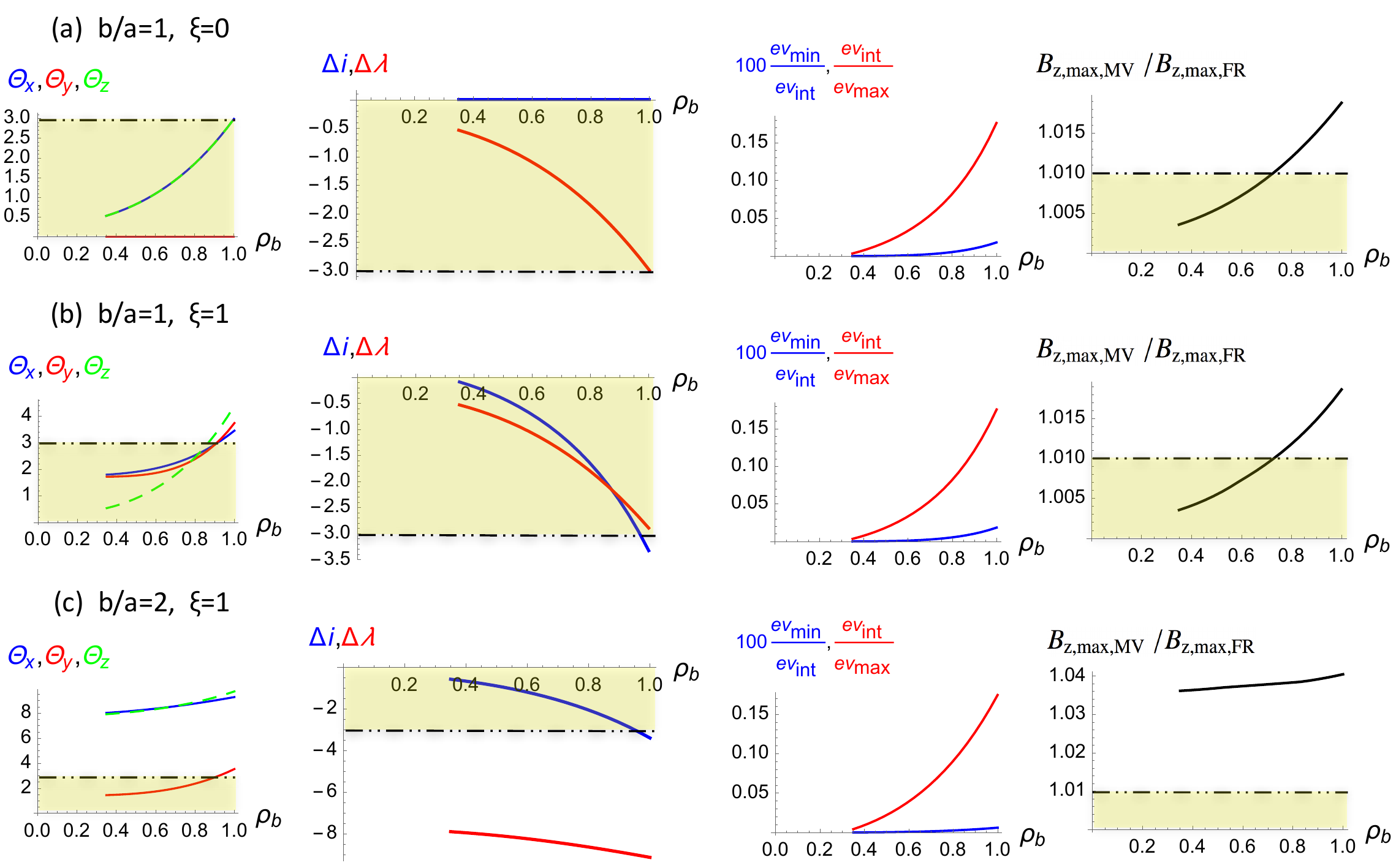}
\caption{Test of \MV{1} applied only to the FR core. The results are shown in function of $\rb = \rF = \rR $, so with flux balance between the front and rear.  
(a,b) $b/a=1$, (c) $b/a=2$. 
(a) $\zeta = 0$ (no expansion), (b,c) $\zeta = 1$ (typical observed expansion). 
 The curves are drawn for $\rb >p$ (otherwise the FR core is not crossed) where $p=0.3$.  \commonCap\ 
The horizontal dashed dotted lines and the yellow regions are landmarks. 
}
\label{fig_rhob}
\end{figure*}

\subsection{Boundaries selected with flux conservation}
  \label{sect_Boundary_Flux}

 Since $\div \vB=0$, the same amount of azimuthal magnetic flux should be present in the FR front and rear \citep[\eg ,][]{Dasso06}. This implies that in the FR frame the same amount of flux crosses the y-z plane before and after the time of the closest spacecraft approach to the FR axis, then for a locally straight FR axis:  
    \BE \label{eq_ByFlux}
    \iint_{\rm FR} \By \  \rmd x \  \rmd z = 0 \,.
    \EE 
    
However, the FR axis of observed MC is expected to be bend toward the Sun. Therefore, following field lines from the front to the rear part, the same amount of flux is located in a slightly longer region along the axis in the front region than at the rear.  Nevertheless, the curvature radius of the axis, $R_{\rm c}$, is large, typically several AUs \citep[\eg ,][]{Janvier15}, compared to the distance $r$ to the FR axis ($r \lesssim 0.1$~AU), then the correction in $r/R_{\rm c}$, is dominant near the border of the FR, but it is still small.
Moreover, the magnetic torque balance is expected to distribute equally the twist along the FR, at least locally. Then, the hypothesis of local invariance by translation along the FR axis is expected to be a good approximation, at least away from the FR legs \citep[see][]{Owens12}. 
 
Within the above approximations, the accumulated magnetic flux between $\tF$ and $\tR$, in the FR frame and per unit length along the axial direction, is
    \BE \label{eq_Byaccumul}
    \frac{\rmd F_{\rm y}(\tF,\tR)}{\rmd z} = \int_{\tF}^{\tR} \By (t') \ \Vx (t') \  \rmd t' \,.
    \EE
Since $\rmd F_{\rm y}/ \rmd z$ includes the observed velocity component $\Vx $ this flux computation includes automatically the details of the local FR expansion.  It is also valid for any FR cross-section shape.
Next, for a given boundary time, either $\tF$ or $\tR$, $\rmd F_{\rm y}(\tF,\tR)/\rmd z = 0$ defines the other FR boundary by imposing the flux balance (as it should be the case for a FR).  In case of multiple solutions, this implies that more than one FR, or a FR and another structure, are present.  Then, the FR, selected by the initial choice of boundary, is defined by minimizing $|\tF-\tR|$ so selecting only one FR.  

Another equivalent formulation is to remark that $\div \vB=0$ implies for a configuration invariant in $z$:
    \BE \label{eq_Binvariant}
    \vB= (\partial A / \partial y, -\partial A / \partial x, ~\Bz (x,y) )\,,
    \EE
where $A(x,y)$ is the $z$ component of vector potential. Integrating any field line implies that it is located on a surface $A(x,y)=$ constant. This property relate any location in the FR front to its corresponding rear position.  Along the spacecraft trajectory, so $y=\yp$ (defined in \eq{spacecraft}), $A(x,\yp)$ can be derived by integration of $\By$.   Next, taking the origin of $A$ at the front boundary this implies $A(t,\yp)= -\rmd F_{\rm y}(\tF,t)/\rmd z$ where the $x$ coordinate has been replaced by time for marking the position. 
Then, \eq{Byaccumul} can also be expressed with the vector potential component $A$, as done, for example, in the Grad-Shrafranov equation for magnetostatic field.

The above property of flux conservation was used to define the back region of MCs \citep[the region which was part of the FR before reconnection at the front occurred,][]{Dasso06,Dasso07} and, more generally, the amount of magnetic flux reconnected with the background field \citep{Ruffenach15}.  Below, we further propose that \eq{Byaccumul} can also be used in a broader context to define coherent sub-FRs within the core of the physical FR.  The aim is to test the stability and to limit the bias of the FR axis determined by the MV variance.  
  
\subsection{Defining the flux rope core}
  \label{sect_Boundary_Sub-FR}
We have shown in \sect{Tests_B}, for FRs with cylindrical symmetry, that the MV bias decreases as the azimuthal field component $\Bt$ or $\Bt /|\vB|$ (\MV{0} or \MV{1}) has a more linear dependance with the distance to the FR axis.  For non singular distributions of the electric current and cylindrically symmetric FRs, $\Bt$ vanishes on the axis.  A Taylor expansion of $\Bt (r)$ implies that $\Bt (r)$ is linear for sufficiently small $r$ values, so the bias of MV is expected to be lower when MV is applied to the core of the FR.  
However, restricting the core has the disadvantage of considering less data points, and for real FRs it implies including relatively more small structures and noise/fluctuations that can compete with the FR structure and produce an additional undesired bias.
Then, we use \eq{Byaccumul} to define sub-regions corresponding to inner FRs in the simulated data.  For the models considered, \eq{Byaccumul} is simply solved by setting $\rF = \rR = \rb$.
When real observations are analyzed, getting the boundaries for this flux-balanced inner sub-FR requires to know its orientation (i.e., $B_y$ in the FR frame is needed to compute $F_y$). 
Thus, it is needed to use a method that feedbacks on the best estimations of the orientation while checking that the flux balance criterion is met.

Next, we describe how to impose the flux balance with data.  The following method was tested with the above models where the flux balance is rigorously defined with $\rF = \rR $.
The method has two main steps: fixing either the front or rear boundary and scanning the other boundary.  More precisely, the method first tests if there is more azimuthal flux at the rear than at the front of the MC, then, if it is the case it defines the FR rear time. If no flux balance is achieved in the previous step, the end time is then used as the rear boundary and the method defines the FR front time when the azimuthal flux is conserved.

We next describe this method with more details.
Let us suppose first that the front boundary is fixed (\eg , it can be defined by a sharp jump in the magnetic and plasma data, or it can be also located inside the MC to study the FR core).  The MV method is recurrently applied to the data limited in between $\tF$ and the scanned values of $\tR$, then the flux balance is computed from \eq{Byaccumul}, in the MV frame (function of $\tR$).  The lowest value of $\tR$ which satisfies the flux balance is selected as the FR rear boundary.  If no flux balance is achieved, the same method is applied by fixing $\tR$ at a sharp jump in the magnetic and plasma data (at the latest time compatible with MC properties), or earlier on, again to study only the FR core. Then, $\tF$ is scanned.  Finally, the largest value of $\tF$ which satisfies the flux balance is retained. 

In summary, present method defines the FR, or its core, which is present at the spacecraft crossing by imposing the azimuthal flux balance. Since the MV eigenvectors are computed with flux balance, the bias on the deduced FR axis orientation is minimized.   Finally, this method of scanning boundaries to impose flux balance can be implemented in other methods, such as fitting a flux rope model. The limitation is the computation time, as a non-linear least-square fit is required for each tested boundary.  Still, the computation time can be limited by including in the method a more efficient search of the zero of a function than a simple scan (the simplest being bisection or Newton's root finding).

\subsection{Results with selecting the flux rope core}
  \label{sect_Results_Sub-FR}
  
As expected, the results of \fig{rhob}a show that the \MV{1} bias is significantly reduced as a smaller central part of the Lundquist's FR is considered up to the limit where the FR is not crossed ($\rb = p$).  This contrast with the eigenvalues being closer as $\rb$ decreases, further showing another example where the ratios of eigenvalues are not good indicators of the quality of the deduced FR axis.   Finally, as a natural consequence of \MV{1} and FR frames becoming closer with a lower $\rb$ value, the axial field is also better recovered when $\rb$ is decreased (right panel of \fig{rhob}a).  

 The above trends with $\rb$ are also obtained with \MV{0} but still with a larger bias than with \MV{1}.  
 Next, similar results are obtained when the expansion is included with a magnitude as typically observed (\fig{rhob}b). 
 However, this improved axis determination is less effective as the FR is flatter, so as $b/a$ is larger (\fig{rhob}c).  
 Of course in application to observations, perturbations of the FR field and the limitation to too few data points will constrain the decrease of $\rb$ so close to the impact parameter $p$ as in \fig{rhob}.   

\section{Biases due to fluctuations}
\label{sect_Fluctuations}

The effects of fluctuations can separate two groups: \\
- a change of position for the boundary determined by the flux balance,\\
- a change of the variances in the x,y,z directions (with fixed boundaries).\\

Let us call $\by$ the fluctuation in the $y$ direction. The flux balance writes:
    \BE \label{eq_ByaccumulFluc}
    \frac{\rmd F_{\rm y}(\tF,\tR)}{\rmd z} = \int_{\tF}^{\tR} (\By (t')+\by (t')) \ \Vx (t') \  \rmd t' 
                                           = 0\,,
    \EE
which is simpler to write in function of $x$:
    \BE \label{eq_ByaccumulFlucX}
    \int_{x_{F}}^{x_{R}} (\By +\by) \ \rmd x    = 0\,.
    \EE    
Let us fix $\tF$. $\tR$ is changing by $\Delta t$, corresponding to the size $\Delta x$, to satisfy the flux balance with fluctuations.  The change of flux due to the main field $\By$ is about $B_{y,a} \Delta x$ where $B_{y,a}$ is the amplitude of the field near the FR boundary.
The change of flux due to a sinusoidal fluctuation of wavelength $l$ is at most $b_{y,a} 2/\pi$, where $b_{y,a}$ is the amplitude of the fluctuation, which corresponds to the contribution of half wavelength (the others cancel).  Inserting these orders of magnitude in \eq{ByaccumulFlucX}, and diving by the FR radius $b$ to normalize, the magnitude of the boundary change $\Delta x$ satisfies:
    \BE \label{eq_ByaccumulPert}
    \frac{\Delta x}{b}  \lesssim \frac{2}{\pi} \frac{b_{y,a}}{B_{y,a}} \frac{l}{b} \,.
    \EE    
Both the last two fractions are typically $<<1$, say on the order of $0.1$, so $\Delta x/b < 0.01$, which will give a very small bias on the axis (\fig{rhoR}a is for $\Delta x/b=0.1$, so the effect would be much smaller, likely a factor of ten smaller).   In conclusion, fluctuations would change significantly the computed boundary, so the axis determined, unless they are both large in amplitude and with a large wavelength.   As an example, to get the effect shown in \fig{rhoR}a we need $\Delta x/b=0.1$, then with $b_{y,a}/B_{y,a} \approx l/b$ both ratio needs to be  $\approx 0.4$, and this is the most favorable case where half the wavelength perturbation is not compensated (odd number of half wavelength across the FR).

For the second effect, the variance of \eq{sigma} is rewritten with fluctuations as:
 \BA   \label{eq_sigmaFluct}
 \sigma_n^2 &=&       \langle (B_n+b_n-\langle B_n+b_n \rangle)^2 \rangle  \nonumber \\
            &\approx& \langle (B_n+b_n-\langle B_n \rangle)^2 \rangle  \nonumber \\
            &\approx& \langle (B_n-\langle B_n \rangle)^2 \rangle
                   +2~\langle (B_n-\langle B_n \rangle)~b_n \rangle
                    + \langle  b_n^2 \rangle  \nonumber \\
            &\approx& \langle (B_n-\langle B_n \rangle)^2 \rangle
                    + \langle  b_n^2 \rangle 
 \EA
 where the linear terms in $b_n$ are neglected because they contribute only by at most half wavelength (same as in previous paragraph).   Then, the contributions of the fluctuations is mostly to increase the variances of all B components.   If isotropic fluctuations are introduced they have no effect in the determination of the variance extremums so on the eigenvectors.  This is not the case for fluctuations orthogonal to $\vB$ (Alfvenic fluctuations are expected in a low $\beta$ plasma).   Still, could the eigenvectors be significantly rotated?   Let suppose a rotation of $x,y$ directions to $x',y'$ in the FR frame. Reducing $\langle  b_n^2 \rangle$ in the $x'$ direction would need a significant correlation of the perturbation in the $x,y$ directions (see Appendix \ref{sect_MV_sym}).  Moreover, this rotation would significantly increase $\langle (B_n-\langle B_n \rangle)^2 \rangle$ by mixing $\By$ with $\Bx$ counteracting the above hypothetical decrease of variance due to fluctuations. 
The same consideration can be done with a rotation from $x,z$ directions to $x',z'$, with the difference that it is less costly to mix $x$ and $z$ component for the variance as they have more similar behavior than compared with the $y$ component.

 In conclusion, fluctuations cannot change significantly the $\iA$ angle. The effect is expected to be larger on $\lA$ angle but still weak unless the fluctuations have a long wavelength, comparable to the FR radius and are of large amplitude.   This is very difficult to achieve as we already know that the effect of the expansion, which also can be thought as a kind of large-scale perturbation, is small.

\section{Conclusion} 
\label{sect_Conclusion}

An accurate determination of the flux rope (FR) axis direction from the interplanetary data of $\vB$ is important  for statistical studies determining the global axis shape, for the determination of the FR physical properties (\eg , the internal distribution of the magnetic twist, the amount of magnetic flux and helicity) as well as for comparing the MC global properties and its orientation itself to the ones determined from its solar source region.  
MV is one method to estimate the direction of the FR axis.
However, as any present method, MV is known to have biases.  Our main aim was to identify them and to correct them as much as possible.    

The orientation of the axis is better defined by the location angle $\lA$ and the inclination $\iA$, as defined in \fig{schema}, rather than the traditional latitude and longitude of the axis, using $\uzGSE$ as the polar angle.   
For example, the biases of the MV method are of different origins and magnitudes for $\lA$ and $\iA$ while the bias origins are mixed for latitude and longitude.  
The angles $\lA$ and $\iA$ are then important to identify, then to correct, the MV biases.   For that, we test MV results with several methods and models.

The eigenvalues ratio provided from applying MV to the FR observations quantifies how different/similar are the variances in the three different directions. The eigenvalues should be sufficiently separated to identify accurately the eigenvectors found by MV. 
However, this eigenvalues ratio is not a proxy for quantifying the accuracy of the estimation of the real FR axis direction, in contrast to the claim of several previous publications.

MV was first applied to the simulated $\vB$ time series computed with a Lundquist's model and its generalisation to an elliptical cross-section.
This introduces an increasing bias on the axis determination as the simulated spacecraft crosses the FR further away from its axis. This bias is reduced when the MV is applied to $\vB/B$ time series, confirming the previous results of \citet{Gulisano07}.  By analyzing cylindrical models we identify the origin of this bias as due to the departure of a linear dependence of the azimuthal field components with the distance to the FR axis.   Including expansion in the modeled FR, with an expansion rate as it is typically observed, does not have a significant effect on the orientation obtained from MV. 

Another bias is due to the selection of the time interval used to apply MV.  In MC observations there are frequently several sets of plausible boundaries depending on which magnetic and plasma data are used to define abrupt change of temporal behavior.
We verify on modeled synthetic FR that the selection of the time interval boundaries is crucial to minimize the bias in the determined FR orientation.  We next solve this difficulty by imposing that the same amount of azimuthal magnetic flux should be measured before and after the closest approach of the FR axis. 
The estimation of the flux balance is computed in the MV frame scanning alternately the rear and the front boundary.  We find, as expected, that the MV bias for the estimation of the FR axis direction is minimized when the azimuthal flux balance is satisfied.   This flux balance technique can be applied to other methods, such as FR model fitting. We also anticipate a decrease in the axis orientation bias since the FR models have an intrinsic flux balance, so that the selected part of the data should also satisfy this constraint.   

The above constraint of flux conservation also allows us to explore a variable part of the FR core as input to MV.  Indeed, the test made with cylindrical models shows that the bias in the axis orientation is lower when an inner part of the FR is selected. 
This effect is weaker for FRs with flatter cross sections. 
In applications to observations, the limitation to the FR core is further justified by the expected presence of larger perturbations at the FR periphery, due to interactions with the external ambient solar wind.

However in observations, the presence of $\vB$ fluctuations implies that the analyzed time interval should be sufficiently large and does not affect the axis determination too much.   Then, a compromise should be taken between too small and too large time intervals.  This compromise is MC case dependent since it is linked to the intensity and distribution of $\vB$ perturbations. This can be realized in studied MCs by analyzing the evolution of $\iA$ and $\lA$ in function of the position of one boundary (the other being computed by flux balance).

  In the present study we limit the application of the MV to MCs with approximately symmetric $B(t)$ observed profiles. This corresponds to about 70\% 
the MCs observed at 1 AU with an asymmetry parameter below 15\% \citep[Figure 3h of][]{Lepping18} or with a distortion parameter (DiP) of $0.5 \pm 0.07$ \citep[Table 4 of][]{Nieves18}. 
  Next, on top of the expansion, which introduces only a weak asymmetry, an intrinsic asymmetry exist especially for the faster MCs \citep{Masias-Meza16}.  This spatial intrinsic asymmetry introduces another bias in the axis determined by the MV.  
  Its correction needs both the development of the MV method, and also the development of asymmetric models to test the amount of remaining bias.  This will be the object of another study.
  
Still, the present results are important for applications to mostly slow MCs.  
For FR close to cylindrical symmetry and in expansion, MV applied to $\vB/B$ time series combined with the flux conservation gives results where the remaining biases are mostly present on the location angle $\lA$ with a magnitude typically proportional to the impact parameter.   It implies that, with a known axis shape, there is a bias on how far the spacecraft crossing occurred from the axis apex, see \citet{Janvier15}. The sign of the bias depends on the sign of the product between the impact parameter and of the magnetic helicity. 

At the opposite, the inclination $\iA$ is well recovered (with a bias less than $4\degree$ for an impact parameter below $0.7$). 
This implies that the direction of the FR axis projected orthogonally to the radial direction (from the Sun) is well determined.  Then, the application of the MV method to $\vB/B$ time series, called \MV{1}, is recommended for studying the amount of global rotation of the FR during its transit from the Sun to the \insitu\ observation position.  
   
\begin{acknowledgements}
 We thank the referee for her/his comments which improved the manuscript.
S.D. acknowledges partial support from the Argentinian grants UBACyT (UBA), PICT-2013-1462 (FONCyT-ANPCyT), and PIDDEF 2014/8 (Ministerio de Defensa, Argentina).
This work was partially supported by a one-month invitation of P.D. to the Instituto de Astronom\'ia y F\'isica del Espacio, by a one-month invitation of S.D. to the Institut d'Astrophysique Spatiale, and by a one-month invitation of S.D. to the Observatoire de Paris.
S.D. is member of the Carrera del Investigador Cien\-t\'\i fi\-co, CONICET.
\end{acknowledgements}

%
\bibliographystyle{aa}
\bibliography{mc}
\IfFileExists{\jobname.bbl}{}
{\typeout{}
\typeout{****************************************************}
\typeout{****************************************************}
\typeout{** Please run "bibtex \jobname" to optain}
\typeout{** the bibliography and then re-run LaTeX}
\typeout{** twice to fix the references!}
\typeout{****************************************************}
\typeout{****************************************************}
\typeout{}
}

\begin{appendix} 

\section{Minimum Variance and Symmetries}
\label{sect_MV_sym}

The mean quadratic deviation, or variance, of the magnetic field $\vB$ in a given direction defined by the unit vector $\un$ is:
 \BE   \label{eq_Varn}
 \Varn = \sigma_n^2 = \frac{1}{N}\sum_{k=1}^{N}
   \left( (\vB^k -\langle \vB \rangle) \cdot \un \right) ^2
 \EE
where the summation is done on the $N$ data points of the time series and $\langle \vB \rangle$ is the mean magnetic field.

Let us apply \eq{Varn} to the FR magnetic components observed (or simulated) by a spacecraft, with $\Bx (t), \By (t), \Bz (t)$ the components in the FR frame (defined in Sect. \ref{sect_Geom_Axis}).
This defines a series of $N$ data points $B_i^k$ ($i=x,y,z$).  
We call the variances of these field components $\Vari$.
We define also the correlations between two specific components, as 
  \BE   \label{eq_Coryz}
  \Corij = \frac{1}{N}\sum_{k=1}^{N}
     (\Bi^k -\langle \Bi \rangle) \; (\Bj^k -\langle \Bj \rangle)
  \EE

\subsection{Is the FR $y$-direction well recovered by MV?}

Let us suppose we are in the FR frame, and let us consider a 2D rotation by an angle $\delta$ around a fixed $\bf{x}$ axis,
so supposing $\uxMV=\uxFR$, in order to analyze if MV will mix the components  $\By$ with $\Bz$ when this rotation is permitted. 
Thus, the new component $B_{y'}$ in the rotated frame ($\bf{y'}$) will be
  \BE   \label{eq_Rotx}
  B_{y'} = \By \cosd - \Bz \sind
  \EE
Then, the variance in the new rotated $\bf{y'}$ axis, $\Varyn (\delta)$, can be written as 
 \BA   
 \Varyn (\delta) 
 &=& \Vary (\cosd)^2 + \Varz (\sind)^2 -2 \sind \cosd \; \Coryz  \nonumber\\
 &=& \frac{\Vary-\Varz}{2} \costd - \Coryz \sintd + \frac{\Vary+\Varz}{2} \nonumber
 \EA
This is of the form $a \cos 2 (\delta-\delta_b) +b$ with $a,b$ two constants with $a>0$ if $\Vary > \Varz$. $\delta_b$ is the bias angle of the MV: the rotation around the axis $\uxMV=\uxFR$ needed to maximize $\Varyn (\delta)$. More precisely:
 \BA   
 a &=& \sqrt{\left( \frac{\Vary-\Varz}{2} \right)^2 + \Coryz^2 } \nonumber\\
 \tan 2 \delta_b &=& -\frac{2\;\Coryz}{\Vary-\Varz} \label{eq_biasy}
 \EA
  
A typical order of magnitude of $\delta_b$ can be estimated with the field $\By = \Bo \sin t ,\Bz = \Bo \cos t $ with $t$ in the interval $[-\pi/2,\pi/2]$ (this keeps the global shape of the $\By$ and $\Bz$ profiles, in particular their symmetry, and it is sufficient for an order of magnitude estimation).  This implies $\Vary \approx  \Bo^2/2$, $\Varz \approx  \Bo^2/10$ and $\Coryz = 0$.   Indeed, in practice we have often in MCs the ordering:
  \BE   \label{eq_orderingyz}
  |\Coryz | << \Varz << \Vary
  \EE
It implies 
  \BE   \label{eq_biasyApprox}
  \delta_b \approx \Coryz/\Vary << 1
  \EE

A similar analysis can be done with $x$ replacing $z$, so considering a rotation by an angle $\delta'$ around the z axis, so supposing $\uzMV=\uzFR$, and thus mixing $\Bx$ with $\By$.  
As above, we have often in MCs the ordering:
  \BE   \label{eq_orderingxy}
  |\Corxy | << \Varx << \Vary
  \EE
This implies
  \BE   \label{eq_biasyApprox2}
  \delta'_b \approx -\Corxy/\Vary << 1
  \EE
which is even better satisfied than \eq{biasyApprox} in MCs when $\Bx$ is lower than $\Bz$, so for a low impact parameter. 
 
 We conclude that the $y$-direction, both orthogonal to the FR axis and spacecraft trajectory, is expected to be well recovered by MV when flux balance is achieved.  This is summarized by a low expected bias for the $\iA$ angle.
 
\subsection{Is the FR axis direction well recovered by MV?}
  The same analysis done in the previous sub-section can be applied to the mix between  $\Bx$ with $\Bz$.  
Here it is clearer to look at the bias of the $x$ axis, as a minimum of variance is expected nearby while a saddle point is expected around the $z$ axis (variance decreases when changing in some directions, but increases in other directions).   This is not a problem as the variance matrix is symmetric, then it has orthogonal eigenvectors, so when the bias on both $x$- and $y$-axis are known, the bias on the $z$-axis can be deduced.
  
   Let us consider a rotation by an angle $\delta_y$ around the y axis,
so supposing $\uyMV=\uyFR$.
  \BE   \label{eq_Roty}
  B_{x'} = \Bx \cosdy - \Bz \sindy
  \EE
  
Then, $\Varxn (\delta_y)$ is written as
 \BA   
 &&\Varxn (\delta_y)  \nonumber\\
 &=& \Varx (\cosdy)^2 + \Varz (\sindy)^2 -2 \sindy \cosdy \; \Corxz  \nonumber\\
 &=& -\frac{\Varz-\Varx}{2} \costdy - \Corxz \sintdy + \frac{\Varx+\Varz}{2} \nonumber
 \EA
just as in previous subsection with $y$ changed to $x$ and extracting a negative sign in front of $\costdy$ to write $\Varxn (\delta_y)$ as $-a \cos 2 (\delta_y-\delta_{y,b}) +b$ so that a minimum of $\Varxn (\delta_y)$ is present close to the $x$-axis for small $\delta_{y,b}$ values 
(with $a,b$ two constants with $a>0$ if $\Varz > \Varx$). $\delta_{y,b}$ is the bias angle of the MV for this rotation: the rotation around the axis $\uyMV=\uyFR$ needs to minimize $\Varxn (\delta_y)$. More precisely:
 \BA   
 a &=& \sqrt{\left( \frac{\Varz-\Varx}{2} \right)^2 + \Corxz^2 } \nonumber\\
 \tan 2 \delta_{y,b} &=& \frac{2\;\Corxz}{\Varz-\Varx} \label{eq_biasyb}
 \EA

Unlike previously, Eqs.~(\ref{eq_orderingyz}) and (\ref{eq_orderingxy}), the ordering   
  \BE   \label{eq_orderingxz}
  |\Corxz | << \Varx << \Varz
  \EE
is typically much less satisfied in MCs, so the bias of the rotation $\delta_{y,b}$
around the $y$-axis, as provided by the minimum of variance, could be important,
and it increases with $\Corxz$ so with the impact parameter. Since $\vec{\uxMV}$ and $\vec{\uzMV}$ are orthogonal this conclusion applies also to the axis direction estimated by MV.

 We conclude that the FR axis, or $z$, direction is expected to have more bias than the $y$ direction, except when the impact parameter is small (which implies a small $\Bx$ component, then a small $\Corxz$).  This is summarized by a larger bias for the $\lA$ angle than for $\iA$ except for small $p$ values.
 
\end{appendix}

\end{document}

%% file: mv1_flux_balance_definitions.tex
\definecolor{ao(english)}{rgb}{0.0, 0.5, 0.0} 

\newcommand{\pc}[1]{{\color{Green}{(P: #1)}}}

\newcommand{\BE}{\begin{equation}}
\newcommand{\EE}{\end{equation}}
\newcommand{\BA}{\begin{eqnarray}}
\newcommand{\EA}{\end{eqnarray}}
 \newcommand{\fig}[1]{Fig.~\ref{fig_#1}}

 \newcommand{\sect}[1]{Sect.~\ref{sect_#1}}
 
 \newcommand{\eq}[1]{Eq.~(\ref{eq_#1})}
 \newcommand{\eqs}[2]{Eqs.~(\ref{eq_#1}) and (\ref{eq_#2})}

\newcommand{\eg}{e.g.}
\newcommand{\etal}{{\it et al.}}
\newcommand{\ie}{i.e.}
\newcommand{\insitu}{in situ}

\newcommand{\degree}{\ensuremath{^\circ}}
\renewcommand{\div}[1]{ {\bf \nabla }. #1 }

\newcommand{\rmd}{{\rm d }}

\newcommand{\uvec}[1]{\hat{\boldsymbol{#1}}}

\renewcommand{\ao}{a_{0}}
\newcommand{\bo}{b_{0}}
\newcommand{\Bo}{B_{0}}
\newcommand{\vB}{\vec{B}}
\newcommand{\vBFR}{\vec{B}_{\rm FR}}
\newcommand{\vBobs}{\vec{B}_{\rm obs}}
\newcommand{\Bt}{B_{\rm \theta}}
\newcommand{\Bx}{B_{\rm x}}
\newcommand{\By}{B_{\rm y}}
\newcommand{\Bi}{B_{\rm i}}
\newcommand{\Bj}{B_{\rm j}}
\newcommand{\by}{b_{\rm y}}
\newcommand{\Bz}{B_{\rm z}}
\newcommand{\BxFR}{B_{\rm x,FR}}
\newcommand{\ByFR}{B_{\rm y,FR}}
\newcommand{\BzFR}{B_{\rm z,FR}}
\newcommand{\BtFR}{B_{\rm \theta,FR}}
\newcommand{\BxMV}{B_{\rm x,MV}}
\newcommand{\ByMV}{B_{\rm y,MV}}
\newcommand{\BzMV}{B_{\rm z,MV}}
\newcommand{\Bzratio}{B_{\rm z,max,MV}~/B_{\rm z,max,FR}}
\newcommand{\diA}{\Delta i}  
\newcommand{\dlA}{\Delta \lambda} 
\newcommand{\Do}{D_{0}}
\newcommand{\ud}{\uvec{d}}
\newcommand{\evint}{ev_{\rm int}} 
\newcommand{\evmax}{ev_{\rm max}} 
\newcommand{\evmin}{ev_{\rm min}} 
\newcommand{\iA}{i} 
\newcommand{\iFR}{i_{\rm FR}} 
\newcommand{\iMV}{i_{\rm MV}} 
\newcommand{\kms}{km.s$^{-1}$}
\newcommand{\lA}{\lambda}
\newcommand{\lFR}{\lambda_{\rm FR}} 
\newcommand{\lMV}{\lambda_{\rm MV}} 
\newcommand{\MV}[1]{MV${#1}$}
\newcommand{\pA}{\varphi_{\rm axis}}

\newcommand{\rb}{\rho_{\rm b}}
\newcommand{\rF}{\rho_{\rm front}}
\newcommand{\rObs}{\rho_{\rm obs}}
\newcommand{\rR}{\rho_{\rm rear}}

\newcommand{\tA}{\theta_{\rm axis}} 
\newcommand{\tGH}{T} 
\renewcommand{\to}{t_{0}}
\newcommand{\tx}{\theta_{\rm x}}
\newcommand{\ty}{\theta_{\rm y}}
\newcommand{\tz}{\theta_{\rm z}}
\newcommand{\tF}{t_{\rm F}}
\newcommand{\tR}{t_{\rm R}}
\newcommand{\un}{\uvec{n}}
\newcommand{\Vc}{V_{\rm c}}

\newcommand{\Vx}{V_{\rm x}}

\newcommand{\uxGSE}{\uvec{x}_{\rm GSE}}
\newcommand{\uyGSE}{\uvec{y}_{\rm GSE}}
\newcommand{\uzGSE}{\uvec{z}_{\rm GSE}}
\newcommand{\uxFR}{\uvec{x}_{\rm FR}}
\newcommand{\uyFR}{\uvec{y}_{\rm FR}}
\newcommand{\uzFR}{\uvec{z}_{\rm FR}}
\newcommand{\uxMV}{\uvec{x}_{\rm MV}}
\newcommand{\uyMV}{\uvec{y}_{\rm MV}}
\newcommand{\uzMV}{\uvec{z}_{\rm MV}}
\newcommand{\yp}{y_{\rm p}}

\newcommand{\cosd}{\cos \delta}
\newcommand{\cosdy}{\cos \delta_y}
\newcommand{\costd}{\cos 2 \delta}
\newcommand{\costdy}{\cos 2 \delta_y}
\newcommand{\sind}{\sin \delta}
\newcommand{\sindy}{\sin \delta_y}
\newcommand{\sintd}{\sin 2 \delta}
\newcommand{\sintdy}{\sin 2 \delta_y}
\newcommand{\Corxy}{{\rm Cor}_{\rm x,y}}
\newcommand{\Corxz}{{\rm Cor}_{\rm x,z}}
\newcommand{\Coryz}{{\rm Cor}_{\rm y,z}}
\newcommand{\Corij}{{\rm Cor}_{\rm i,j}}
\newcommand{\Varn}{{\rm Var}_{\rm n}}
\newcommand{\Varyn}{{\rm Var}_{\rm y'}}
\newcommand{\Varxn}{{\rm Var}_{\rm x'}}
\newcommand{\Varx}{{\rm Var}_{\rm x}}
\newcommand{\Vary}{{\rm Var}_{\rm y}}
\newcommand{\Varz}{{\rm Var}_{\rm z}}
\newcommand{\Vari}{{\rm Var}_{\rm i}}

\newcommand{\commonCap}{
  Left column: $\tx,\ty,\tz$ are the angles in degrees between the x, y, z directions, respectively, of the FR and MV frames.  
  Middle left column: $\diA$ and $\dlA$ quantify the difference of axis orientations between the FR and the one found by MV, so the orientation biases.
  Middle right column: ratios of the eigenvalues $\evmin$, $\evint$ and $\evmax$.
  Right column: ratio of the maximum values of $\BzMV$ to $\BzFR$ (amplified by a factor of 100 for the first ratio).}
  
